\documentclass[%
superscriptaddress,
preprint,
amsmath,amssymb,
prd,
]{revtex4-1}
\usepackage{graphicx}
\usepackage{dcolumn}
\usepackage{bm}
\usepackage[colorlinks]{hyperref}
\usepackage{fancyhdr}
\makeatother

\begin{document}

\title{Dynamics of Vortex Electrons in a Uniform Magnetic Field Across Nonrelativistic and Relativistic Regimes} 

\author{Yujian Lin}
\affiliation{Department of Optics and Optical Engineering, School of Physical Sciences, University of Science and Technology of China, Hefei 230026, China}
\affiliation{State Key Laboratory of Ultra-intense Laser Science and Technology, Shanghai Institute of Optics and Fine Mechanics (SIOM), Chinese Academy of Sciences, Shanghai 201800, China}
\author{Zhigang Bu}
\affiliation{State Key Laboratory of Ultra-intense Laser Science and Technology, Shanghai Institute of Optics and Fine Mechanics (SIOM), Chinese Academy of Sciences, Shanghai 201800, China}
\author{Shiyu Liu}
\affiliation{State Key Laboratory of Ultra-intense Laser Science and Technology, Shanghai Institute of Optics and Fine Mechanics (SIOM), Chinese Academy of Sciences, Shanghai 201800, China}
\author{Taoming Guo}
\affiliation{Department of Optics and Optical Engineering, School of Physical Sciences, University of Science and Technology of China, Hefei 230026, China}
\affiliation{State Key Laboratory of Ultra-intense Laser Science and Technology, Shanghai Institute of Optics and Fine Mechanics (SIOM), Chinese Academy of Sciences, Shanghai 201800, China}

\author{Liangliang Ji}
\email{jill@siom.ac.cn}
\affiliation{State Key Laboratory of Ultra-intense Laser Science and Technology, Shanghai Institute of Optics and Fine Mechanics (SIOM), Chinese Academy of Sciences, Shanghai 201800, China}

\date{\today}

\begin{abstract}
The dynamics of a vortex electron in a transverse uniform magnetic field involves not only the cyclotron motion of the wave packet center but also the evolution of its internal wavefunction structure.
To study this process, we develop a quantum framework and investigate the dynamics
of vortex electrons in both the nonrelativistic and relativistic regimes.
A discrete Bessel-Gaussian initial wave packet is expanded in the
basis of Dirac eigenstates to obtain the time-dependent wavefunction and to analyze its spatiotemporal
evolution.
In the nonrelativistic regime, the results deviate significantly from the rigid point-dipole model, as the intrinsic kinetic orbital angular momentum exhibits dynamics more complex than simple Larmor precession, owing to the nonstationary internal current distribution of the wave packet.
In the relativistic regime, the nonlinear dependence of the energy levels on the quantum numbers destroys the equal spacing of the spectrum.
Consequently, the relative phases among the eigenstate components of the wave packet evolve nonuniformly, resulting in wave-packet spreading and deformation, accompanied by modified intrinsic orbital angular momentum dynamics.
We introduce a dimensionless parameter to quantify this relativistic dephasing, providing an approximate criterion for selecting physical parameters and evolution times to mitigate the resulting wave-packet distortion.
These results clarify the limitations of the rigid point-dipole model, highlight the essential role of internal wavefunction structure, and provide theoretical guidance for understanding and manipulating relativistic vortex electrons.
\end{abstract}
\maketitle

\section{\label{sec:Introduction} Introduction}

Vortex beams, as a unique type of structured wave,
are characterized by a helical wavefront structure,
a phase singularity line of zero intensity, and an
intrinsic orbital angular momentum (OAM) along the
propagation direction~\cite{bliokh2017theory}.
The concept of vortex photons was first proposed by
Allen et al. thirty years ago~\cite{allen1992orbital},
laying the foundation for exploring wave fields carrying OAM.
Subsequently, experimental evidence for
vortex electrons was obtained~\cite{uchida2010generation}, prompting
extensive studies of vortex matter waves.
Learning the dynamical behavior of vortex electrons
in external electromagnetic fields thus becomes critical.  ~\cite{bliokh2007semiclassical,silenko2017manipulating,silenko2018relativistic,karlovets2026angular,gallatin2012propagation,bliokh2012electron,guzzinati2013observation,greenshields2014angular,han2017classical,babiker2015electron}.
It is well-known that spin precesses in external fields following the Thomas-BMT equation~\cite{thomas1927kinematics,bargmann1959precession}. A natural question is whether the intrinsic OAM of a vortex electron exhibits an analogous evolution. A common starting point for addressing this question is the quasi-classical description, in which the dynamics of a vortex wave packet is approximated by the evolution of its expectation values, allowing the packet to be modeled as a point charge carrying a rigid magnetic dipole associated with its intrinsic OAM. Within this framework, an early semiclassical description of electron wave packets carrying phase vortices revealed Berry-curvature-induced corrections to their dynamics and the associated OAM Hall effect~\cite{bliokh2007semiclassical}. A subsequent relativistic treatment in the Foldy-Wouthuysen representation, applicable to arbitrary electromagnetic fields, predicted a radiative orbital polarization effect analogous to the Sokolov-Ternov effect~\cite{silenko2018relativistic}. More recently, these ideas have been extended to accelerator physics, including analyses of OAM preservation, radiative losses, and resonance effects during the acceleration of vortex particles to high energies~\cite{karlovets2026angular}.

Despite its convenience, the quasi-classical approach generally cannot fully resolve the internal structure and spatiotemporal evolution of realistic vortex wave packets. A more complete description therefore requires a wavefunction-based framework. Within this framework, the evolution of nonrelativistic vortex electron wave packets in uniform magnetic fields has been studied using the exact path-integral propagator, as reflected in the spatial probability density~\cite{gallatin2012propagation}.
In the relativistic regime, wavefunction-based descriptions of vortex electrons in uniform magnetic fields have already been developed~\cite{van2017nonuniform,rajabi2017relativistic}. Exact Dirac solutions in the form of Laguerre-Gaussian beams have been used to analyze probability and current distributions, angular momentum expectation values, and intrinsic spin-orbit coupling~\cite{rajabi2017relativistic}. Related exact vortex solutions reveal nonuniform azimuthal currents and spin polarization, as well as the absence of spin-orbit mixing in selected Landau-Zeeman ground states~\cite{van2017nonuniform}. Relativistic paraxial treatments have further described beam-width oscillations in uniform longitudinal magnetic fields~\cite{zou2021general} and breathing dynamics in time-dependent longitudinal fields~\cite{wang2021breathing}. These studies establish important properties of relativistic vortex electrons in fields parallel to the beam propagation axis, but do not directly address the time-dependent dynamics of localized vortex wave packets in transverse magnetic fields. Understanding these dynamics is important for characterizing the evolution of vortex structure and intrinsic OAM during magnetic bending, and for assessing their preservation in beam manipulation, transport, and storage-ring applications.

When the average momentum and intrinsic OAM of a vortex electron are initially perpendicular to a uniform magnetic field, a key question is whether its OAM follows the simple Larmor precession predicted by a rigid point-dipole model, and how the internal dynamics of the wave packet affect this evolution. A single stationary vortex Landau state aligned with the field does not describe the initial transverse vortex packet; its evolution must instead be constructed from a coherent superposition of magnetic-field eigenstates. A wavefunction-based treatment can therefore resolve the wave-packet deformation and interference between eigenstate components that are absent from the rigid point-dipole description.

To address this question, we develop a fully quantum theoretical framework to
investigate the spatiotemporal evolution of vortex
electrons in a uniform magnetic field. We project the incident initial vortex state onto the Dirac eigenstates in a perpendicular magnetic field and construct the time-dependent wave function from the projection amplitudes, thereby describing the dynamical behavior. In the nonrelativistic regime, our description of the probability density is consistent with the existing theoretical results \cite{gallatin2012propagation}. However, the calculated evolution of the expectation values of the OAM differs from the simple Larmor precession predicted by the rigid point-dipole model. Extending to the relativistic regime, we find that the nonequidistant energy spectrum destroys the phase synchrony among eigenstate components, leading to nontrivial wave-packet spreading and modifying the evolution of the intrinsic OAM. This analysis offers a wavefunction-based perspective on relativistic vortex electron dynamics in magnetic fields and provides new insight into how internal wave-packet structure  and relativistic spectral effects jointly shape the evolution of vortex electron wave packets.

The paper is organized as follows. Sec.~\ref{sec:cbde} 
presents a complete set of eigenstates of the Dirac Hamiltonian in a uniform magnetic field.
Sec.~\ref{sec:init_state} introduces a special initial vortex state suited for constructing the time-dependent function. 
In Sec.~\ref{sec:projection_amplitude}, the initial vortex state is expanded in the Dirac eigenstate basis 
to obtain the projection amplitudes and
construct the time-dependent wavefunction. 
Sec.~\ref{sec:nrel_ve} examines the results in the
nonrelativistic regime, investigates the dynamics of
the intrinsic OAM, and highlights its marked
difference from the rigid point-dipole model. 
Sec.~\ref{sec:rel_ve} extends the analysis to the
relativistic regime and discusses the mechanism of
wave packet deformation induced by the nonequidistant relativistic energy levels. 
Finally, Sec.~\ref{sec:conclusion} summarizes the paper.

\section{\label{app:theoretical_framework} Theoretical framework}
\subsection{\label{sec:cbde} Dirac Eigenstates in a Uniform Magnetic Field}
We consider the Dirac equation in a uniform magnetic field using natural units $c=\hbar=1$:
\begin{equation}
\label{eq:dirac}
(\mathrm{i} \partial_t -\hat{H}) \psi(\boldsymbol{x},t) =0 ,
\end{equation}
here, the Hamiltonian is given by
$\hat{H}=\boldsymbol{\alpha} \cdot (\boldsymbol{p}+e\boldsymbol{A}) +\beta m_e $.
The magnetic field $\boldsymbol{B}=B \bar{z}$ is chosen to point along the positive $z$ direction, and the vector potential is taken in the symmetric gauge, 
$\boldsymbol{A}= -\frac{B y}{2} \bar{x}+\frac{B x}{2} \bar{y}$,
where $e=|e|>0$ is the elementary charge
(with $\bar{x}$, $\bar{y}$, and $\bar{z}$ denoting
the unit vectors in the corresponding directions).

The Dirac equation can be solved by following the methods
in Refs.~\cite{van2017nonuniform,rajabi2017relativistic},
to obtain stationary solutions in cylindrical coordinates
(for the detailed solution process, see Appendix~\ref{app:solve_dirac}),
where the positive-energy part is given by:

\begin{equation}
\label{eq:psi_pos_energy}
\begin{aligned}
j>0:\  \psi^{+}_{nlsp_z}(\boldsymbol{x})=C_E\,\mathrm{e}^{\mathrm{i}p_zz}\left[\delta_{s,\frac{1}{2}} 
\begin{pmatrix}
\mathcal{J}_n^l(r,\varphi)\\
0\\
\frac{p_z}{E+m_e} \mathcal{J}_n^l(r,\varphi)\\
\mathrm{i}\frac{\sqrt{2eB}}{E+m_e}\mathcal{J}_n^{l+2s}(r,\varphi)
\end{pmatrix}+
\delta_{s,-\frac{1}{2}}
\begin{pmatrix}
0\\
\mathcal{J}_n^l(r,\varphi)\\
-\mathrm{i}\frac{\sqrt{2eB}}{E+m_e}N_{nls}\mathcal{J}_n^{l+2s}(r,\varphi)\\
-\frac{p_z}{E+m_e} \mathcal{J}_n^l(r,\varphi)
\end{pmatrix}\right],
\\ 
j<0:\  \psi^{+}_{nlsp_z}(\boldsymbol{x})=C_E \,\mathrm{e}^{\mathrm{i}p_zz} \left[\delta_{s,\frac{1}{2}}
\begin{pmatrix}
\mathcal{J}_n^l(r,\varphi)\\
0\\
\frac{p_z}{E+m_e} \mathcal{J}_n^l(r,\varphi)\\
-\mathrm{i}\frac{\sqrt{2eB}}{E+m_e}N_{nls}\mathcal{J}_{n+2s}^{l+2s}(r,\varphi)
\end{pmatrix}+
\delta_{s,-\frac{1}{2}}
\begin{pmatrix}
0\\
\mathcal{J}_n^l(r,\varphi)\\
\mathrm{i}\frac{\sqrt{2eB}}{E+m_e}\mathcal{J}_{n+2s}^{l+2s}(r,\varphi)\\
-\frac{p_z}{E+m_e} \mathcal{J}_n^l(r,\varphi)
\end{pmatrix}\right].
\end{aligned}
\end{equation}

The negative-energy part is given by
$\psi^{-}_{nlsp_z}(\boldsymbol{x})=\gamma^5\gamma^0\,\psi^{+}_{nlsp_z}(\boldsymbol{x})$, where $\gamma^\mu$ are the Dirac gamma matrices and $\gamma^5=\mathrm{i}\gamma^0\gamma^1\gamma^2\gamma^3$.
For the eigenstate $\Psi_{n l s p_z}^\pm$, the superscript $\pm$
corresponds to the positive and negative energy solutions
of the Dirac equation, while the subscripts $n, l, s, p_z$
label the radial, orbital angular momentum, spin quantum numbers,
and the momentum in the $z$ direction, respectively.
The total angular momentum quantum number is $j=l+s$.
The normalization coefficient $C_E$ and the energy $E$
are defined respectively as:
$C_E=\frac{1}{2\pi}\sqrt{\frac{eB(E+m_e)}{2 E}}\sqrt{\frac{n!}{(n+|l|)!}}$,
$E=\epsilon_{nlsp_z}=\sqrt{p_z^2+m_e^2+2eBN_{nls}}$, where
$N_{nls}=n+\frac{l+|l|+1}{2}+s$.
Furthermore, the term $2eB N_{nls}=\epsilon_Z^2+\epsilon_L^2$,
with $\epsilon_Z^2=2eBs,\  \epsilon_L^2=2eB\left(n+\frac{l+|l|+1}{2}\right)$.
Here, $\epsilon_Z$ is the interaction energy of the electron's spin magnetic moment (Zeeman energy), and $\epsilon_L$ is the sum of the electron's orbital kinetic energy and the interaction energy of its orbital magnetic moment (Landau energy).
In addition, we introduce the function
$\mathcal{J}^l_n(r,\varphi)=e^{\mathrm{i}l\varphi}e^{\frac{-\tilde{r}^2}{2}}\tilde{r}^{|l|}L^{|l|}_n(\tilde{r}^2)$
and the dimensionless radius $\tilde{r}=\sqrt{\frac{eBr^2}{2}}$.

For an electron in a uniform magnetic field, the time-dependent wavefunction
can be expanded in the eigenstate basis by projecting onto each eigenstate~\cite{greiner1990relativistic}. Accordingly, an arbitrary time-dependent wavefunction $\Psi(\boldsymbol{x},t)$ can be written as:

\begin{equation}
\label{eq:mtdfw}
\begin{aligned}
\Psi(\boldsymbol{x},t)&=\sum\limits_{n,l,s} \int_{-\infty}^{\infty} \mathrm{d} p_z \, \langle \boldsymbol{x} |\mathrm{e}^{-\mathrm{i} \hat{H}t} \left(|\psi^{+}_{nlsp_z}\rangle\langle\psi^{+}_{nlsp_z}|+|\psi^{-}_{nlsp_z}\rangle\langle\psi^{-}_{nlsp_z}|\right) |\Psi_i \rangle   \\
&=\sum\limits_{n,l,s} \int_{-\infty}^{\infty} \mathrm{d} p_z \left[C^+_{nlsp_z} \psi^+_{nlsp_z}(\boldsymbol{x})\ \mathrm{e}^{- \mathrm{i}\epsilon_{nlsp_z}t}+C^-_{nlsp_z} \psi^-_{nlsp_z}(\boldsymbol{x})\ \mathrm{e}^{ \mathrm{i}\epsilon_{nlsp_z}t}\right],  
\end{aligned}
\end{equation}
where
\begin{equation}
\label{eq:pm_C_coef}
C^{\pm}_{nlsp_z}=\int \mathrm{d} ^3x'\psi^{\pm \dagger}_{nlsp_z}(\boldsymbol{x}')\Psi_i(\boldsymbol{x}'),
\end{equation}
is the amplitude of the initial state $\Psi_i(\boldsymbol{x})$
projected onto an eigenstate $\psi^{\pm}_{nlsp_z}(\boldsymbol{x})$.

Here we neglect radiation effects and electron-positron pair production. In general, discarding
negative-energy states breaks the completeness of the
basis such that the time-dependent wavefunction at the
initial time $\Psi^+(\boldsymbol{x},0)$ cannot strictly
reconstruct the original initial state $\Psi_i(\boldsymbol{x})$.
Nevertheless, for the parameters considered here, the
negative-energy weight is much smaller than the
positive-energy one (the projected probability onto the positive-energy states exceeds that onto the negative-energy states by more than five orders of magnitude). The truncated wavefunction
thus preserves the main physical features of the
vortex electron and is sufficient for our purposes.
Accordingly, we consider the positive-energy contribution in the time-dependent wavefunction:

\begin{equation}
\label{eq:mtdfw_puls}
\Psi^+(\boldsymbol{x},t)=\sum\limits_{n,l,s} \int_{-\infty}^{\infty} \mathrm{d}p_z \, C^+_{nlsp_z} \psi^+_{nlsp_z}(\boldsymbol{x}) e^{-\mathrm{i}\epsilon_{nlsp_z}t}.
\end{equation}

\subsection{\label{sec:init_state}Choice of Initial State}
To derive the projection amplitude
$C^+_{nlsp_z}$ analytically, the initial state should
capture the main physical characteristics of a vortex electron (in particular, the phase singularity line of zero intensity, the helical phase structure, and the expectation value of OAM) and remain tractable for the projection
calculation. For commonly used vortex electron
states, the projection integrals onto the complete
basis are usually not available in analytic form. We
therefore adopt the discrete electron Bessel state
proposed in Ref.~\cite{wong2024free} as a simplified
initial state. As shown later, this state retains the main features of a vortex electron and also allows an
analytic evaluation of the projection amplitude. It
is given by

\begin{equation}
\label{eq:discrete_bessel}
\Psi_i(\boldsymbol{x})=\frac{1}{\sqrt{N_p}}\sum\limits_{m=0}^{N_p-1}  \ \mathrm{e}^{\mathrm{i} j_{0}\phi_m}\ u(\theta_i,\phi_m)\ \mathrm{e}^{\mathrm{i}\boldsymbol{p_m} \cdot \boldsymbol{x}},
\end{equation}
where $\phi_m = 2m\pi/{N_p}$, $N_p$ denotes the total number of plane-wave components in the initial state, $\theta_i$ represents the opening angle  (polar angle) and $\phi_m$ is the azimuthal angle of the $m$-th plane wave, and $j_0$ is the order of the Bessel state.
A larger $N_p$ yields a better approximation to the continuous
Bessel state, but at a higher computational cost.
It is known that Bessel state is nonlocal and 
thus not suitable for our purpose of studying the
evolution of a localized vortex electron wave packet
in a magnetic field. We therefore attach a Gaussian
wave packet to this discrete Bessel state
\cite{vaity2015perfect} (An overall normalization factor is omitted here and below; all probability densities and expectation values are normalized by the norm of the constructed time-dependent wavefunction.):

\begin{equation}
\label{eq:discrete_bessel_gauss}
\Psi_i(\boldsymbol{x})= \frac{1}{\sqrt{N_p}}\sum\limits_{m=0}^{N_p-1}  \ \mathrm{e}^{\mathrm{i}j_{0}\phi_m}\ u^{s_y^\uparrow}(\theta_i,\phi_m)\ \mathrm{e}^{\mathrm{i}\boldsymbol{p_m} \cdot \boldsymbol{x}}\ \mathrm{e}^{-\frac{1}{2\sigma^2}(x^2+y^2+z^2)}.
\end{equation}

With the initial wave packet propagating along the positive $y$ axis
(perpendicular to the magnetic field as depicted in Fig.~\ref{fig:3D_and_phase_schematic}(a)),
$\phi_m$ and $\theta_i$ denote the azimuthal and polar angles regarding the $y$ axis.
Consequently, the discrete momentum is expressed as:

\begin{figure}[htbp]
    \centering
    \includegraphics[width=1\textwidth]{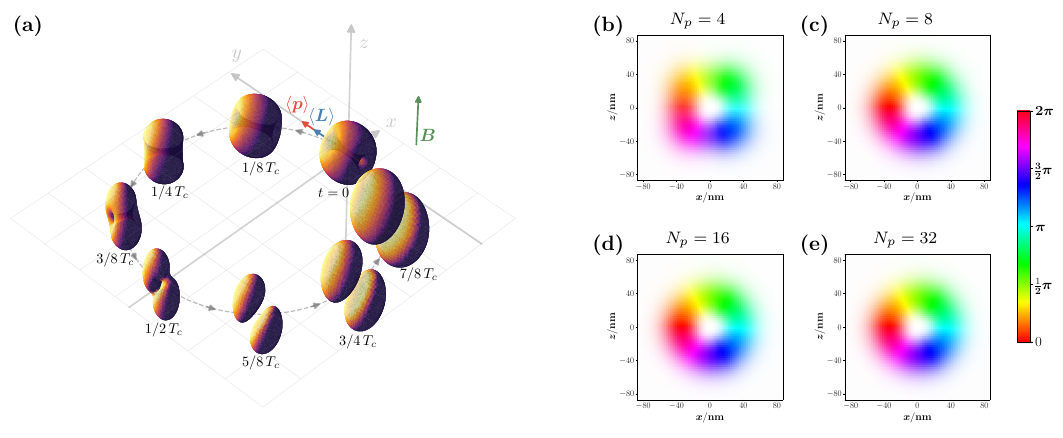}
    \caption{(a) Schematic illustration of the evolution of a vortex electron in a uniform magnetic field $B\parallel \bar{z}$. At the initial time $t=0$, both the mean momentum $\langle \boldsymbol{p} \rangle$ and the intrinsic OAM $\langle \boldsymbol{L} \rangle$ are aligned along the $y$-axis. The figure shows the cyclotron trajectory and the wave-packet deformation at representative times within one cyclotron period $T_c$. The parameters are $B=2\,\mathrm{T}$, $p_{my}=0.3\,\mathrm{keV}$, $p_{m\perp}=6\,\mathrm{eV}$, $\sigma=40\,\mathrm{nm}$, $l_0=1$, and $N_p=32$. (b)–(e) Phase distributions in the $zx$ cross section at $t=0$, corresponding to the initial wave packet shown in panel (a), for different total numbers of plane-wave components: (b) $N_p=4$, (c) $N_p=8$, (d) $N_p=16$, and (e) $N_p=32$. The color bar denotes the phase from $0$ to $2\pi$.}
    \label{fig:3D_and_phase_schematic}
\end{figure}

\begin{equation}
\label{eq:p_m}
\begin{aligned}
\boldsymbol{p}_m&=p_{m\perp}  \sin\phi_m\bar{x}+p_{my}\bar{y}+p_{m\perp}  \cos\phi_m\bar{z}
\\&=p_{my}(\tan\theta_i\sin\phi_m\bar{x}+1\bar{y} +\tan\theta_i\cos\phi_m\bar{z}), 
\end{aligned}
\end{equation}
here, $p_{my}$ and $p_{m\perp}$ denotes the longitudinal and transverse momentum components with respect to the propagation direction ($y$-axis), satisfying $\tan\theta_i=p_{m\perp}/p_{my}$.
Furthermore, the bispinor $u^{s_y^\uparrow}(\theta_i,\phi_m)$ is chosen as
the positive helicity eigenstate with the $y$-axis as the quantization axis,
taking the specific form:

\begin{equation}
\label{eq:bispinor_y}
u^{s_y^\uparrow}(\theta_i,\phi_m) =\begin{pmatrix}
\sqrt{(E_0+m_e)}\ \omega^{\uparrow}_{y,1}\\
\sqrt{(E_0+m_e)}\  \omega^{\uparrow}_{y,2}\\
\sqrt{(E_0-m_e)} \ \omega^{\uparrow}_{y,1}\\
\sqrt{(E_0-m_e)}\  \omega^{\uparrow}_{y,2}
\end{pmatrix},
\end{equation}
here,
$E_0=\sqrt{\boldsymbol{p}_m^2+m_e^2}$, $\omega^{\uparrow}_{y,1} =\frac{1}{\sqrt{2}}(\mathrm{e}^{-\frac{\mathrm{i}}{2}\phi_m}\cos\frac{\theta_i}{2}+\mathrm{e}^{\frac{\mathrm{i}}{2}\phi_m}\sin\frac{\theta_i}{2})$, and $\omega^{\uparrow}_{y,2} =\frac{\mathrm{i}}{\sqrt{2}}(\mathrm{e}^{-\frac{\mathrm{i}}{2}\phi_m}\cos\frac{\theta_i}{2} -\mathrm{e}^{\frac{\mathrm{i}}{2}\phi_m}\sin\frac{\theta_i}{2})$. Since the initial bispinor is chosen to be a positive helicity state, the intrinsic OAM $l_0$ of the initial vortex wave packet and the total angular momentum (i.e., the order of the Bessel state) $j_0$ satisfy the relation
$l_0=j_0-\frac{1}{2}+\mathcal{O}(\theta_i^2)$ in the paraxial limit.

Based on the above definition of the initial state Eq.~\ref{eq:discrete_bessel_gauss}, the time-dependent wavefunction can be conveniently obtained from Eq.~\ref{eq:mtdfw_puls}.  Fig. 1(a) shows the trajectory of a vortex electron wave packet in a uniform magnetic field $B$ along the $z$-axis, together with the time evolution of its probability-density isosurfaces. At $t=0$, both the initial momentum $\langle \boldsymbol{p}\rangle$ and the intrinsic OAM $\langle \boldsymbol{L}\rangle$ are oriented along the $y$-axis.
Here, the cyclotron period is defined as $T_c=\frac{2\pi}{eB}\gamma m_e
=\frac{2\pi}{eB}\sqrt{p_{my}^2+m_e^2}$.
Figures 1(b)–(e) show the phase distributions on the $z-x$ cross section at $t=0$, corresponding to the initial wave packet shown in Fig. 1(a), with different total numbers of plane-wave components of $N_p=4$, 8, 16, 32, respectively. Although the negative-energy components are neglected and discretized Bessel–Gaussian states are used, the time-dependent wavefunction at $t=0$ retains the basic characteristics of a vortex wave packet, including a phase singularity
line of zero intensity and a helical phase structure. 
As shown in Figs. 1(b)–(e), the phase distribution gradually approaches a continuous vortex pattern with increasing $N_p$. The variation becomes much less pronounced for $N_p>8$, and the phase structure converges at $N_p=32$. Therefore, $N_p=32$ is adopted throughout the calculations in this work. The expectation value of the OAM is presented later in Fig.~\ref{fig:L_total_NR}, where at the initial time $\langle L_y\rangle$ matches the prescribed value $l_0$.

\subsection{\label{sec:projection_amplitude}Projection amplitude}

Using Eq.~\ref{eq:pm_C_coef}, together with Eqs.~\ref{eq:psi_pos_energy} and \ref{eq:discrete_bessel_gauss},
the projection amplitude $C^\pm_{nlsp_z}$ can be calculated.
The integration in Eq.~\ref{eq:pm_C_coef} is decomposed into the following parts:

The part independent of coordinates:
\begin{equation}
\label{eq:I_0}
I_0(n,l,s,p_z,m)=\frac{\mathrm{e}^{\mathrm{i}j_{0}\phi_m}}{2 \pi \sqrt{N_p}}\sqrt{\frac{eB \left(\epsilon_{nlsp_z}+m_e\right)}{2\epsilon_{nlsp_z}}}\sqrt{\frac{n!}{(n+|l|)!}}.
\end{equation}

The part dependent on $z$:
\begin{equation}
\label{eq:I_z}
I_z(p_z,m)=\int_{-\infty}^{\infty}\mathrm{d}z\ \mathrm{e}^{i(p_{mz}-p_z)z} \mathrm{e}^{-\frac{z^2}{2\sigma^2}}=\sqrt{2\pi\sigma^2}\ \mathrm{e}^{-\frac{\sigma^2}{2}(p_{mz}-p_z)^2}. 
\end{equation}

The $xy$-dependent part, shown here for the representative case $j>0$ and $s>0$ (the other cases differing only by sign changes and index shifts), is
\begin{equation}
\begin{aligned}
I_{xy}^{\pm}(n,l,s,p_z,m)
=&(\sqrt{E_0\pm m_e}\pm \frac{ p_z\sqrt{E_0\mp m_e}}{\epsilon_{nlsp_z} + m_e})\omega^{\uparrow}_{y,1}Q_{nlm}\\
&\mp \mathrm{i}\frac{ \sqrt{ 2 e B(E_0\mp m_e)}}{\epsilon_{nlsp_z} + m_e} \omega^{\uparrow}_{y,2} Q_{n(l+1) m}.
\end{aligned}
\end{equation}

The remaining three cases and the evaluation of $Q_{nlm}$ are detailed in Appendix~\ref{app:I_xy_and_Q_nlm}. Therefore, the projection amplitudes are given by:
\begin{equation}
\label{eq:pm_C_coef_final}
C^{\pm}_{nlsp_z}=\sum_{m=0}^{N_p -1} I_0(n,l,s,p_z,m) I_z(p_z,m) I_{xy}^{\pm}(n,l,s,p_z,m).
\end{equation}

Substituting Eq.~\ref{eq:pm_C_coef_final} into Eq.~\ref{eq:mtdfw_puls} yields the time-evolution function
for the initial state $\Psi_i(\boldsymbol{x})$. 
In the numerical evaluation of Eq.~\ref{eq:mtdfw_puls}, the summation and integration are truncated to finite ranges.
With suitably chosen truncation ranges, 
the time-dependent wavefunction at the initial time reproduces the prescribed initial vortex state with good accuracy, 
enabling the subsequent dynamical analysis. Details of the ranges chosen in Eq.~\ref{eq:mtdfw_puls} 
are provided in Appendix~\ref{app:choose_range}.
 
\section{\label{sec:nrel_ve}The Nonrelativistic Regime}
\begin{figure}[htbp]
    \centering
    \includegraphics[width=1\textwidth]{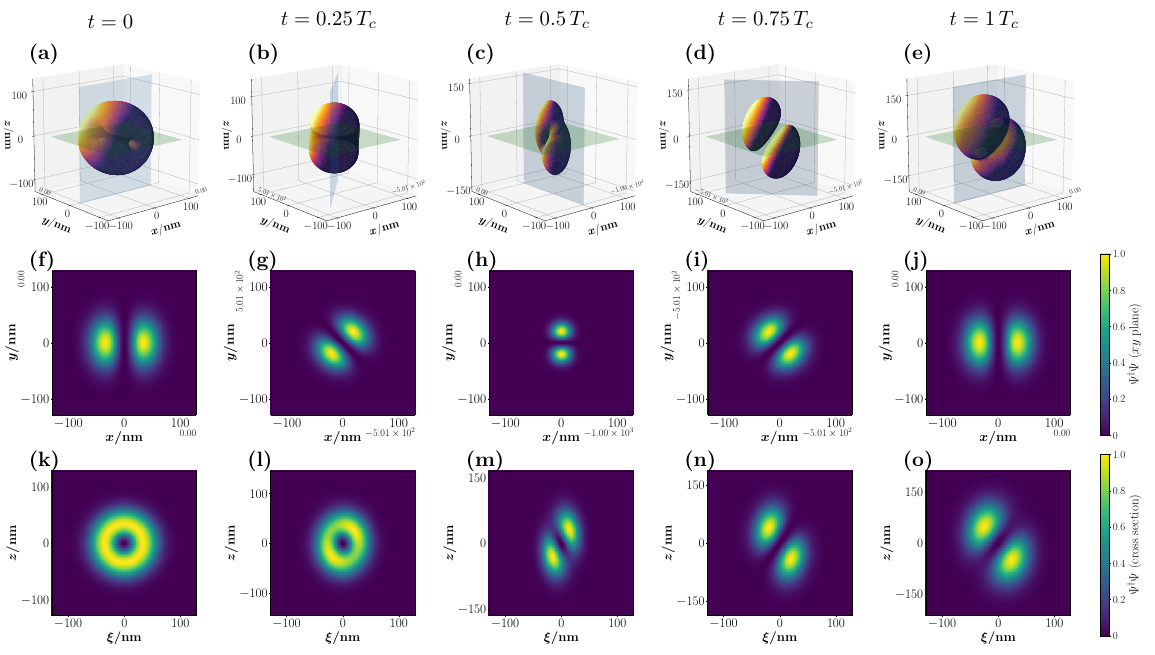}
    \caption{
Probability density distributions of the vortex electron wave packet
($B=2\, \mathrm{T},\ p_{my}=0.3 \,\mathrm{keV},\ p_{m\perp}=6\,\mathrm{eV},\ \sigma= 40\, \mathrm{nm},\ l_0=1$) at five time instants (columns):
Panels (a)–(e) display 3D probability-density isosurfaces centered at the wave packet centroid. The green plane is the $xy$ plane at $z=0$. The blue plane is a cross section perpendicular to both the green plane and the singularity line. Panels (f)–(j) and (k)–(o) show 2D distributions in the green and blue planes, respectively. The definition of the $\xi$ axis is given in Eq.~\ref{eq:bar_xi}. Coordinate offsets are indicated next to the axes.}
    \label{fig:evolution_2}
\end{figure}

Before studying relativistic vortex electrons, we first verify that, in the nonrelativistic regime, our results from the Dirac equation correspond to those from the Schr\"odinger equation~\cite{gallatin2012propagation}.
We consider the following parameters: magnetic field $B=2\,\mathrm{T}$,
initial longitudinal momentum  $p_{my}=0.3\,\mathrm{keV}$,
transverse momentum $p_{m\perp}=6\,\mathrm{eV}$,
wave packet width $\sigma= 40\,\mathrm{nm}$,
and intrinsic OAM $l_0=1$.
Figure~\ref{fig:evolution_2} shows the wave packet
evolution in one cycle. To highlight the change in the wave packet shape,
the coordinate origin at each time is shifted to
the corresponding wave packet centroid. 
On the cross section perpendicular to the $xy$ plane and the
singularity line  (i.e., the vortex core, where the intensity is zero), we define coordinates
$(\xi,z)$, where $\xi$ is the horizontal axis. Its
positive unit vector is defined by
\begin{equation}
\label{eq:bar_xi}
\bar{\xi}=
\begin{cases}\cos \frac{\pi t}{T_c}\bar{x}+\sin \frac{\pi t}{T_c}\bar{y} ,\ \mathrm{for}\ 0 \le t \le \frac{T_c}{2}\\
-\cos \frac{\pi t}{T_c}\bar{x}-\sin \frac{\pi t}{T_c}\bar{y} ,\ \mathrm{for}\ \frac{T_c}{2} <t\le T_c .   
\end{cases}
\end{equation}

The nonrelativistic results derived from the Dirac equation are highly consistent with those from the Schr\"odinger equation, and this agreement is particularly evident in the probability density on the $xy$ plane shown in panels (f)–(j) of Fig.~\ref{fig:evolution_2}, which closely resembles Fig.~3 of Ref.~\cite{gallatin2012propagation}. The evolution can be decomposed into parts parallel and perpendicular to the magnetic field. Along the magnetic-field direction (the $z$ axis), the kinetic term leads to quantum dispersion during free propagation, and the packet therefore broadens along the $z$ direction, as shown by the continuous spreading with time in panels (k)–(o) of Fig.~\ref{fig:evolution_2}. As noted in Ref.~\cite{gallatin2012propagation,kim1973gaussian}, in the $xy$ plane perpendicular to the field, free transverse dispersion competes with the effective two-dimensional harmonic-oscillator confinement (harmonic-oscillator-like focusing) arising from the magnetic-field-related quadratic term $\frac{e^2B^2}{8m_e}(x^2+y^2)$. As a result, the transverse width exhibits periodic breathing oscillations in time, as illustrated in panels (f)–(j) of Fig.~\ref{fig:evolution_2}. In addition, the angular-momentum coupling term $\frac{eB}{2m_e}L_z$ characterizes an overall rotation of the wave-packet envelope relative to its centroid, as illustrated in panels (a)–(e) and (f)–(j) of Fig.~\ref{fig:evolution_2}. The corresponding angular velocity is $\omega_c/2$, where $\omega_c=eB/m_e$ is the cyclotron frequency.

\begin{figure}[htbp]
    \centering
    \includegraphics[width=1\textwidth]{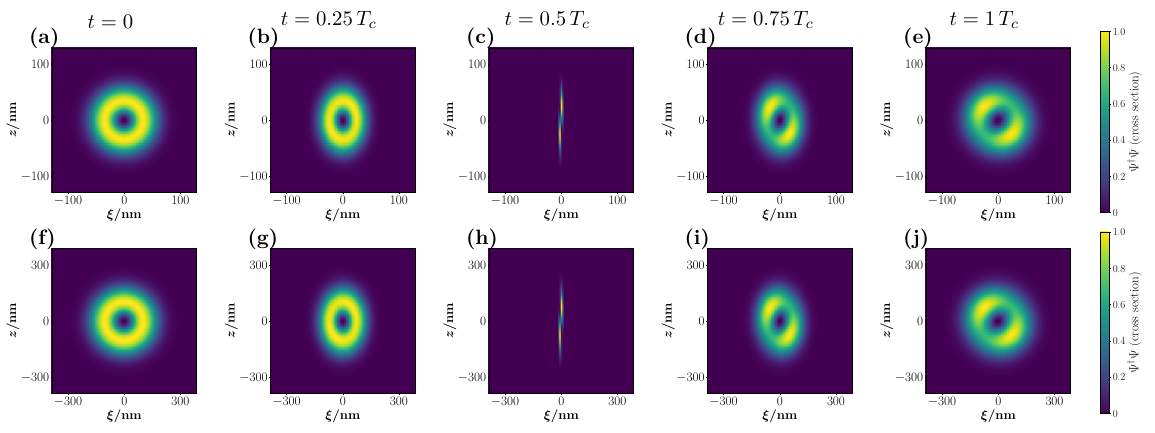}
    \caption{Probability density distributions of the vortex electron wave packet in the cross section (as defined in Fig.~\ref{fig:evolution_2}) at five time instants.
Panels (a)–(e) correspond to $B=18\,\mathrm{T}$, $p_{my}=0.3\,\mathrm{keV}$, $p_{m\perp}=6\,\mathrm{eV}$, $\sigma=40\,\mathrm{nm}$, and $l_0=1$;
panels (f)–(j) correspond to $B=2\,\mathrm{T}$, $p_{my}=0.3\,\mathrm{keV}$, $p_{m\perp}=2\,\mathrm{eV}$, $\sigma=120\,\mathrm{nm}$, and $l_0=1$.}
    \label{fig:evolution_1}
\end{figure}

The evolution shown in panels (k)–(o) of Fig.~\ref{fig:evolution_2} should not be interpreted as a mere free expansion along the $z$ direction. A similar tilted two-lobed structure has also been reported in spatiotemporal optical vortices, where it is attributed to the asymmetric distribution of the transverse energy flow~\cite{bekshaev2024spatiotemporal}. For vortex electrons, the corresponding physical mechanism is the nonstationarity and asymmetry of the internal current distribution of the extended wave packet. This internal current shear deforms the singularity line during the evolution and gives rise to the tilted two-lobed structure in the cross section.

In order to estimate the broadening along the $z$ direction, it is useful to first compare with the free evolution of an ordinary Gaussian electron wave packet, for which the width evolves as $\sigma(t)=\sigma_0\sqrt{1+t^2/(2m_e\sigma_0^2)^2}.$
This expression shows that the broadening is governed by the dimensionless combination $t/(2m_e\sigma_0^2)$. For one cyclotron period, this parameter becomes $T_c/(2m_e\sigma^2)
=\pi\sqrt{p_{my}^2+m_e^2}/(m_e eB\sigma^2)$,
which reduces to $\pi/(eB\sigma^2)$
in the nonrelativistic limit. 
Accordingly, increasing either $\sigma$ or $B$, or both, suppresses the broadening along the $z$ direction. 
In Fig.~\ref{fig:evolution_1}(a)–(e), with the magnetic field increased to $B=18\,\mathrm{T}$, and in Fig.~\ref{fig:evolution_1}(f)–(j), with the wave-packet width increased to $\sigma=120\,\mathrm{nm}$, the probability density exhibits the same qualitative evolution: The initially hollow ring-shaped distribution becomes elliptically deformed at $t=0.25T_c$, then collapses into a strongly compressed stripe-like profile near $t=0.5T_c$. In the second half of the cycle, it re-expands into a tilted two-lobed structure and finally returns approximately to the initial ring-shaped configuration at $t=T_c$. 
Compared with Fig.~\ref{fig:evolution_2}(o), after one period, the present case exhibits a much weaker separation into two lobes. This difference indicates that suppressing the broadening along the $z$ direction helps maintain the compactness of the ring-shaped profile and significantly reduces the tendency toward a pronounced two-lobed splitting.

This structural evolution also raises a subtle question regarding the relation between the singularity line and the intrinsic OAM. The singularity line is determined by the nodal structure of the wave packet and provides a geometrical characterization of the vortex. In the symmetric gauge used here, the intrinsic canonical OAM is closely related to the phase circulation around this singularity line, because the canonical momentum operator directly probes the phase gradient of the wave function. Therefore, although the canonical OAM is gauge dependent and should not be regarded as a gauge-invariant mechanical angular momentum, it can still serve, within the symmetric gauge, as a useful phase-sensitive indicator of the evolution of the vortex structure. In particular, Ref.~\cite{gallatin2012propagation} and panels (f)–(j) of Fig.~\ref{fig:evolution_2} showed that the rotation angle of the singularity line is only half of the momentum rotation angle. One might be tempted to conclude that the intrinsic OAM precesses in the $xy$ plane at the Larmor frequency and reverses its direction after one cyclotron period. Such an interpretation, however, should be treated with caution. It is based primarily on the probability density and therefore does not fully capture the phase structure of the vortex electron. 
To quantify the evolution of the phase structure, we calculate the expectation values of the canonical OAM operators $L_x$ and $L_y$, where $L_x=-\mathrm{i}(y\partial_z-z\partial_y)$ and $L_y=-\mathrm{i}(z\partial_x-x\partial_z)$.
The above canonical OAM is introduced to characterize the phase-sensitive vortex structure. However, for comparison with the rigid point-dipole model, the relevant quantity is the intrinsic kinetic OAM, $\mathbf{L}^{\rm kin}_{\rm int}=\left\langle\left(\mathbf r-\langle\mathbf r\rangle\right)\times\left(\boldsymbol{\pi}-\langle\boldsymbol{\pi}\rangle\right)\right\rangle ,$ where $\boldsymbol{\pi}$ is the kinetic momentum. This quantity is gauge invariant and represents the mechanical angular momentum of the wave packet. We therefore also calculate the intrinsic kinetic OAM and compare its evolution with the prediction of the rigid point-dipole model.

\begin{figure}[htbp]
    \centering
    \includegraphics[width=1\textwidth]{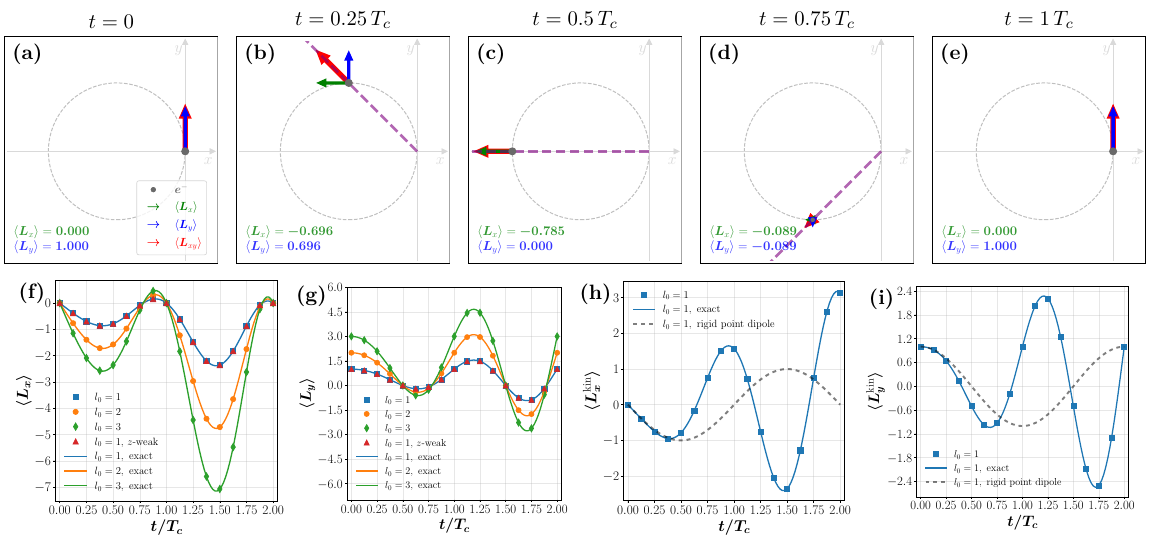}
    \caption{Expectation values of the canonical and kinetic OAM for nonrelativistic vortex electrons. Panels (a)–(e) illustrate the transverse canonical OAM components $\langle \boldsymbol{L}_x \rangle$, $\langle \boldsymbol{L}_y \rangle$, and $\langle \boldsymbol{L}_{xy} \rangle$ along the cyclotron trajectory at different times, using the same parameters as in Fig.~\ref{fig:evolution_2}. Gray dots mark the electron position; green, blue, and red arrows denote $\langle \boldsymbol{L}_x \rangle$, $\langle \boldsymbol{L}_y \rangle$, and $\langle \boldsymbol{L}_{xy} \rangle$ (in units of $\hbar$), respectively. The gray dashed circles indicate the cyclotron trajectory, and the purple dashed lines connect the electron to the origin. Panels (f) and (g) show the corresponding time evolution of the canonical OAM components $\langle L_x \rangle$ and $\langle L_y \rangle$ versus $t/T_{\mathrm{c}}$. The blue, orange, and green symbols correspond to $l_0=1$, 2, and 3, respectively, with the same parameters as in Fig.~\ref{fig:evolution_2} otherwise. The red symbols correspond to $l_0=1$ and weaker broadening along $z$ ($z$-weak), with the other parameters taken from Fig.~\ref{fig:evolution_1}(f)–(j). The blue, orange, and green solid curves show the corresponding exact solutions for $l_0=1,2,$ and $3$, respectively. Panels (h) and (i) show the time evolution of the kinetic OAM components $\langle L_x^{\rm kin}\rangle$ and $\langle L_y^{\rm kin}\rangle$ for $l_0=1$. The symbols denote the results obtained with the same parameters as in Fig.~\ref{fig:evolution_2}, the solid curves show the exact solutions, and dashed curves show the Larmor precession predicted by the rigid point-dipole model.}
    \label{fig:L_total_NR}
\end{figure}

Fig.~\ref{fig:L_total_NR} shows the evolution of the canonical and kinetic OAM of a nonrelativistic vortex electron.
Figures~\ref{fig:L_total_NR}(f) and
\ref{fig:L_total_NR}(g) show
$\langle L_x \rangle$ and
$\langle L_y \rangle$ as functions of the
normalized time $t/T_c$, respectively. The blue,
orange, and green symbols correspond to the nonrelativistic cases with initial angular
momentum $l_0=1$, 2, and 3. 
Analysis of Fig~\ref{fig:L_total_NR}(a)-(g) reveals that these results exhibit
a complex evolutionary behavior:
Regardless of the initial OAM, the magnitude of the total canonical OAM expectation value in the $xy$ plane is not conserved. 
Instead, it undergoes a process of initial decrease followed by an increase.
Simultaneously, the canonical OAM direction flips along the radial direction, i.e., 
it changes from initially pointing radially outward (from the origin to the electron) to pointing radially inward (from the electron to the origin). 
It is the combined effect of this magnitude fluctuation 
and the  direction flip that causes the canonical OAM of the
electron not to reverse after one complete cyclotron period, 
but rather to maintain its initial direction. Nevertheless, the purple dashed lines in Figure~\ref{fig:L_total_NR}(a)-(e) indeed indicate the direction of the singularity line (as shown in Figure~\ref{fig:evolution_2}(f)-(j)); therefore, consistent with the phase-circulation picture discussed above, the intrinsic canonical OAM remains aligned with the singularity line throughout the evolution. Furthermore, the blue and red symbols in Figures~\ref{fig:L_total_NR}(f) and
\ref{fig:L_total_NR}(g) show that the wave-packet deformation associated with the broadening along the $z$-direction has a negligible effect on the expectation value of the canonical OAM.

In panels (h) and (i) of Fig.~\ref{fig:L_total_NR}, the symbols show the kinetic OAM components $\langle L_x^{\rm kin}\rangle$ and $\langle L_y^{\rm kin}\rangle$ for $l_0=1$, obtained using the same parameters as in Fig.~\ref{fig:evolution_2}. The dashed curves show the Larmor precession predicted by the rigid point-dipole model, which treats the vortex electron as a point charge carrying a rigid magnetic dipole induced by its intrinsic OAM. In this model, the OAM vector has a fixed magnitude and precesses uniformly about the magnetic field, giving sinusoidal oscillations with constant amplitude and a period of $2T_c$. The kinetic OAM components initially follow this prediction but later show clear differences. They oscillate more rapidly, and their oscillation amplitudes increase over the time interval shown. These differences indicate that the rigid point-dipole model does not fully describe the kinetic OAM dynamics of the vortex electron.

Comparing the canonical OAM components in panels (f) and (g)
with the kinetic components in panels (h) and (i) of Fig.~\ref{fig:L_total_NR} for $l_0=1$
also reveals distinct time evolution. To understand this
difference, we expand the intrinsic kinetic OAM as
$\mathbf L_{\rm int}^{\rm kin}
=\langle\mathbf r\times\boldsymbol{\pi}\rangle
-\langle\mathbf r\rangle\times\langle\boldsymbol{\pi}\rangle$.
For the wave packet considered here, direct evaluation of
the expectation values gives
$\langle z\rangle=\langle p_z\rangle=0$, consistent with
the classical centroid trajectory confined to the $xy$
plane. Since $\pi_z=p_z$, the second term therefore has
vanishing $x$ and $y$ components. Using the previously
defined symmetric gauge, we obtain
$\langle L_x^{\rm kin}\rangle
=\langle y\pi_z-z\pi_y\rangle
 =\langle L_x\rangle-\frac{eB}{2}\langle xz\rangle,\ 
\langle L_y^{\rm kin}\rangle
=\langle z\pi_x-x\pi_z\rangle
 =\langle L_y\rangle-\frac{eB}{2}\langle yz\rangle$.
Thus, the differences between the canonical and kinetic
OAM components in the $xy$ plane are determined by the
mixed spatial moments $\langle xz\rangle$ and
$\langle yz\rangle$. These moments describe correlations
within the wave packet and need not vanish even when
$\langle z\rangle=0$. A related distinction was discussed by
Bliokh et al.~\cite{bliokh2012electron} for electron vortex states
in magnetic fields. For the OAM component along the field,
the corresponding relation, expressed in our charge
convention, is
$\langle L_z^{\rm kin}\rangle
=\langle L_z\rangle+eB\langle x^2+y^2\rangle/2$.
Greenshields et al.~\cite{greenshields2014angular} further showed
that changes in the mean-square radius can lead to
time-dependent kinetic $L_z$ even when canonical $L_z$
is conserved, with total angular momentum conservation
recovered by including the electromagnetic field.
Here, the corresponding corrections to the OAM components
in the $xy$ plane depend on mixed spatial moments rather
than the mean-square radius. Moreover, rotational symmetry
about the magnetic-field axis does not imply conservation
of canonical $L_x$ or $L_y$, allowing both the canonical
and kinetic components in the $xy$ plane to evolve with time.

Alternatively, we revisit the intrinsic OAM equation of motion for vortex electrons in the Heisenberg picture. As detailed in Appendix~\ref{app:Exact_nonrelativistic_solution}, the nonrelativistic Hamiltonian is quadratic in a uniform magnetic field, so the second-moment equations form a closed system and the transverse intrinsic kinetic OAM can be solved analytically.

Using the exact solution, the analytical evolution formula for the kinetic OAM in the nonrelativistic case is given by Eq.~\ref{eq:exact_solution}, and that for the canonical OAM is given by Eq.~\ref{eq:canonical_oam}. The results obtained from these formulas are shown as the solid curves in Fig.~\ref{fig:L_total_NR}(f)-(i). Moreover, these solid curves are found to agree well with the symbols plotted in Fig.~\ref{fig:L_total_NR}(f)-(i), which represent the numerical results obtained in this paper. This clearly validates the correctness of our numerical method and provides confidence for its application in the relativistic case. In addition, as analyzed in Appendix~\ref{app:Exact_nonrelativistic_solution}, the OAM evolution is determined by the initial second moments of position and kinetic momentum, including their cross correlations, rather than by the detailed functional form of the wave-packet envelope. Therefore, as shown by the red and blue symbols in Fig.~\ref{fig:L_total_NR}(f) and (g), the wave-packet deformation associated with broadening along the $z$-direction has no effect on the OAM evolution.

\begin{figure}[htbp]
    \centering
    \includegraphics[width=0.5\textwidth]{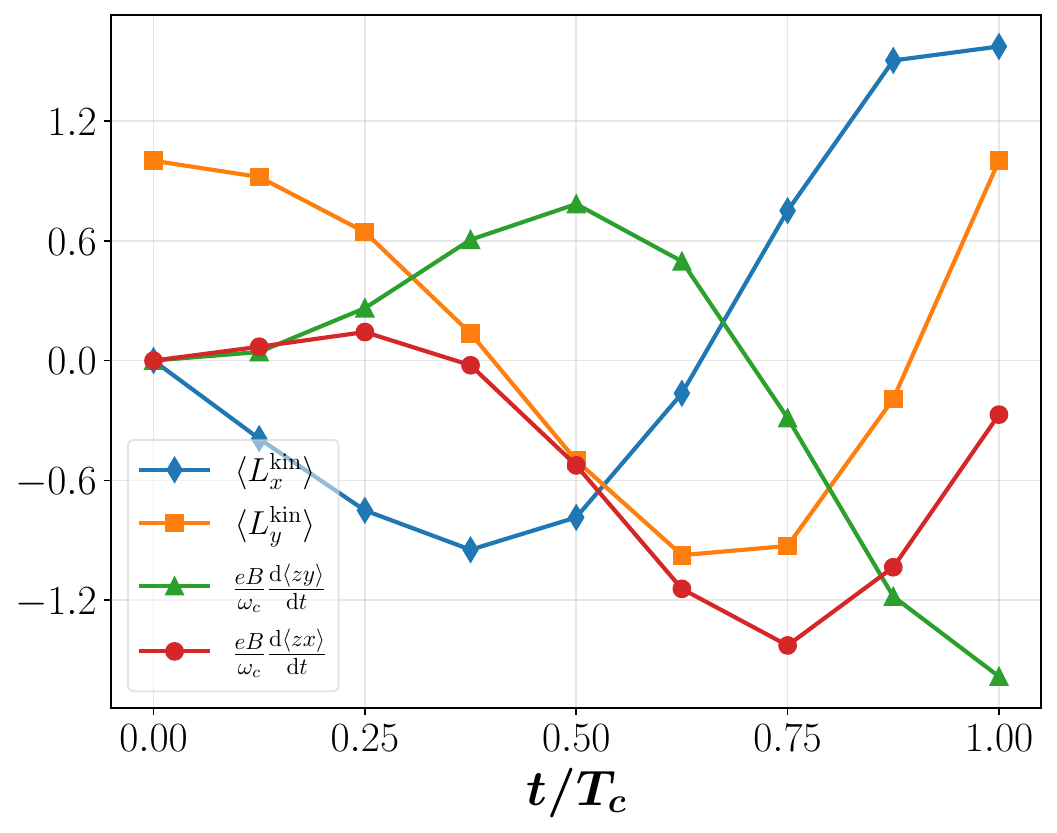}
    \caption{Time evolution of the expectation values $\langle L_x^{\text{kin}} \rangle$, $\langle L_y^{\text{kin}} \rangle$, $\frac{eB}{\omega_c}\frac{\text{d} \langle zy \rangle}{\text{d}t}$, and $\frac{eB}{\omega_c}\frac{\text{d} \langle zx \rangle}{\text{d}t}$ for nonrelativistic vortex electrons. The blue, orange, green, and red solid curves denote $\langle L_x^{\text{kin}} \rangle$, $\langle L_y^{\text{kin}} \rangle$, $\frac{eB}{\omega_c}\frac{\text{d} \langle zy \rangle}{\text{d}t}$, and $\frac{eB}{\omega_c}\frac{\text{d} \langle zx \rangle}{\text{d}t}$, respectively, as functions of the normalized time $t/T_{\mathrm{c}}$. The other parameters are the same as those used in Fig.~\ref{fig:evolution_2}.}
    \label{fig:zxy}
\end{figure}

Beyond validating the numerical results, the exact equations of motion also clarify why the kinetic OAM evolution departs from the Larmor precession predicted by the rigid point-dipole model. As shown in Eq.~\ref{eq:gauge_invariant_equations}, the transverse kinetic OAM components are coupled not only through the usual Larmor-precession terms but also through the time derivatives of the mixed spatial moments, $(eB/2)d\langle zx\rangle/dt$ and $(eB/2)d\langle zy\rangle/dt$. In the absence of these additional contributions, the equations would reduce to those describing pure Larmor precession. To assess their relative importance, Fig.~\ref{fig:zxy} compares $\langle L_x^{\mathrm{kin}}\rangle$ and $\langle L_y^{\mathrm{kin}}\rangle$ with the scaled mixed-moment derivatives $(eB/\omega_c)d\langle zy\rangle/dt$ and $(eB/\omega_c)d\langle zx\rangle/dt$, respectively. The scaling allows a direct comparison with the corresponding kinetic OAM terms in Eq.~\ref{eq:gauge_invariant_equations}. The mixed-moment derivatives, represented by the green and red curves, are relatively small at early times. This explains why the kinetic OAM initially agrees well with the prediction of the rigid point-dipole model in Fig.~\ref{fig:L_total_NR}(h) and (i). However, around $t=T_{\mathrm c}/2$ and thereafter, their magnitudes become comparable to those of the kinetic OAM components. These gauge-invariant quantities characterize the nonstationarity of the internal current distribution, whose evolving spatial correlations modify the OAM dynamics and lead to deviations from pure Larmor precession. Such internal dynamics of an extended wave packet are beyond the description provided by the rigid point-dipole model.

In summary, the vortex electron is essentially a structured wave.
Its vortex properties are strongly correlated with the spatial distribution of the wavefunction,
and its wave nature is inseparable. Therefore, it cannot be treated simply 
as a point charge carrying a rigid magnetic dipole induced by its intrinsic OAM.

\section{\label{sec:rel_ve}The Relativistic Regime}
\begin{figure}[htbp]
    \centering
    \includegraphics[width=1\textwidth]{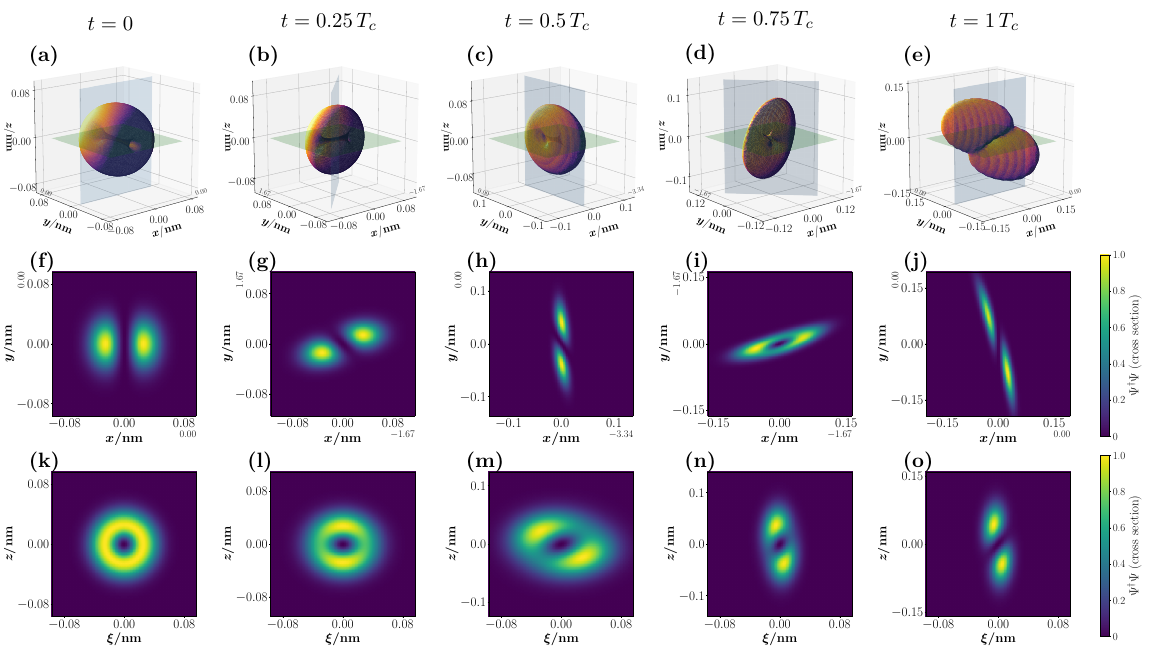}
    \caption{Probability density distributions of the vortex electron wave packet
($B=4 \times 10^6 \, \mathrm{T},\ p_{my}=2\,\mathrm{MeV},\ p_{m\perp}=8 \,\mathrm{keV}, \ \sigma= 30\,\mathrm{pm},\ l_0=1$) at five time instants (columns):
Panels (a)–(e) display 3D probability density isosurfaces centered on the wave packet centroid.
The green plane is the $xy$ plane at $z=0$.
The blue plane is a cross section ($\bar{\xi}$ as defined
in Eq.~\ref{eq:bar_xi}). Panels (f)–(j) and (k)–(o) show 2D distributions in the green and blue planes, respectively.
Coordinate offsets are indicated next to the axes.
}
    \label{fig:evolution_3}
\end{figure}

Turning to the relativistic regime, the initial momentum is set to $p_{my}=2\,\text{MeV}$ to examine the impact of relativistic effects on the evolution and to compare with the nonrelativistic case. 
Before the actual calculations, challenges in numerical implementation must be addressed. 
Under a typical laboratory magnetic field strength $B=1\,\text{T}$, this choice yields $p_Q=\sqrt{\frac{2}{eB}(p_{mx}^2+p_{my}^2)}\sim 10^5$ (see Appendix~\ref{app:choose_range} for more details on $p_Q$). 
Consequently, Laguerre polynomials of order as high as $\frac{p_Q^2}{4}\sim 10^{10}$ must be evaluated. 
At such high orders, conventional iterative algorithms become prohibitively inefficient; 
therefore, an asymptotic expansion of the relevant functions is employed (see Appendix~\ref{app:Laguerre} for details).
Even with asymptotic expansion techniques, the overall computational cost remains substantial
due to the massive number of eigenstates required for superposition.
To illustrate  the 3D structure of relativistic vortex electrons 
and the evolution of the expectation value of the OAM
at an acceptable computational cost, a compromise was made in choosing the physical parameters:
by selecting a strong magnetic field $B=4\times 10^6\,\mathrm{T}$,
the relevant quantity in the calculation is reduced to about $\frac{p_Q^2}{4}\approx 1.69\times 10^4$.
For these parameters (other parameters are
$p_{my}=2\,\mathrm{MeV},\ p_{m\perp}=8 \,\mathrm{keV},\ \sigma= 30\,\mathrm{pm},\ l_0=1$),
the probability density distributions of the vortex electron at different times 
are shown in Fig.~\ref{fig:evolution_3}.

\begin{figure}[htbp]
    \centering
    \includegraphics[width=1\textwidth]{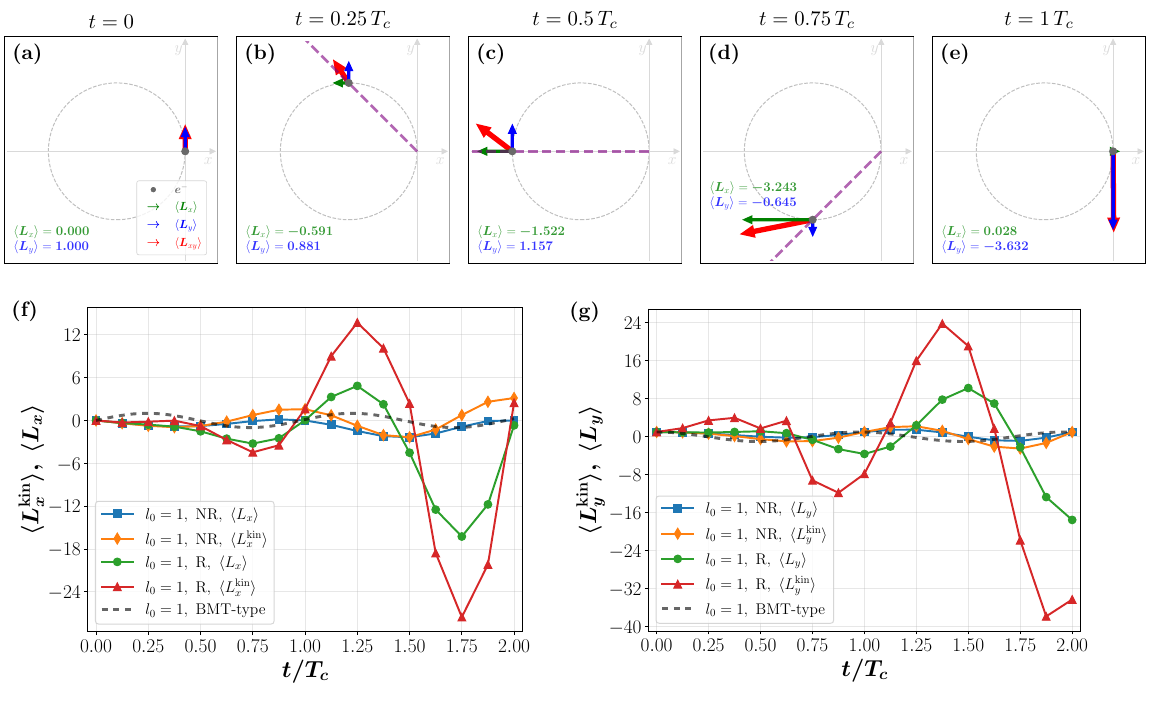}
    \caption{Expectation values of the OAM for relativistic vortex electrons. Panels (a)–(e) illustrate $\langle \boldsymbol{L}_x \rangle$, $\langle \boldsymbol{L}_y \rangle$, and $\langle \boldsymbol{L}_{xy} \rangle$ along the cyclotron trajectory at different times, using the same parameters as in Fig.~\ref{fig:evolution_3}. Gray dots mark the electron position; green, blue, and red arrows denote $\langle \boldsymbol{L}_x \rangle$, $\langle \boldsymbol{L}_y \rangle$, and $\langle \boldsymbol{L}_{xy} \rangle$ (in units of $\hbar$), respectively. The gray dashed circles indicate the cyclotron trajectory, and the purple dashed lines connect the electron to the origin. Panels (f) and (g) show the corresponding time evolution of the canonical and kinetic OAM components along the $x$ and $y$ axes, respectively, versus $t/T_{\mathrm{c}}$. The blue and orange solid curves  show the nonrelativistic (NR) results for the canonical and kinetic OAM, respectively, with the same parameters as in Fig.~\ref{fig:evolution_2}. The green and red solid curves  show the relativistic (R) results for the canonical and kinetic OAM, respectively, with the same parameters as in Fig.~\ref{fig:evolution_3}. The gray dashed curves show the precession predicted by the Thomas-BMT-type equation for $l_0=1$.}
    \label{fig:L_total_R}
\end{figure}

\begin{figure}[htbp]
    \centering
    \includegraphics[width=1\textwidth]{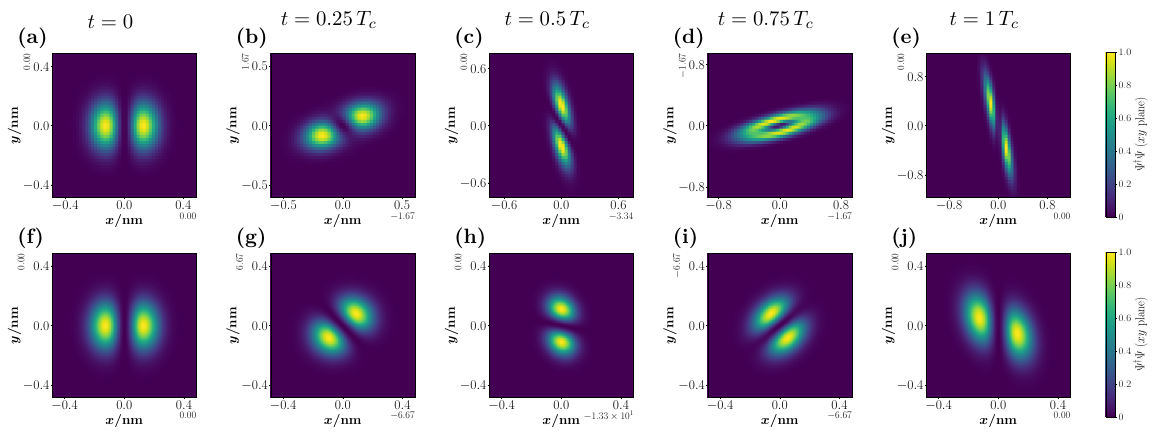}
    \caption{Probability density distributions of the vortex electron wave packet in the
    $xy$ plane ($z=0$)
    at five time instants (columns): Panels (a)–(e) correspond to $B=10^5\,\mathrm{T}$,
    $p_{my}=2\,\mathrm{MeV}$,
    $p_{m\perp}=1.6\,\mathrm{keV}$,
    $\sigma=150\,\mathrm{pm}$, and $l_0=1$; panels (f)–(j) correspond to $B=10^5\,\mathrm{T}$,
    $p_{my}=0.2\,\mathrm{MeV}$,
    $p_{m\perp}=1.6\,\mathrm{keV}$,
    $\sigma=150\,\mathrm{pm}$, and $l_0=1$ (coordinate offsets are indicated).}
    \label{fig:evolution_4}
\end{figure}

Furthermore, to investigate the evolutionary behavior under a weaker magnetic field,
we set the magnetic field to $B=10^5\,\mathrm{T}$ (for which $\frac{p_Q^2}{4}=3.38\times10^5$).
Under these conditions (with other parameters being
$p_{my}=2\,\mathrm{MeV},\  p_{m\perp}=1.6 \, \mathrm{keV},\ \sigma= 150\,\mathrm{pm},\ l_0=1$),
the obtained 2D probability density distributions of the electron
are shown in Fig.~\ref{fig:evolution_4}.

Comparing Figs.~\ref{fig:evolution_2} and \ref{fig:evolution_3}, we find that,
significantly differing from the nonrelativistic case, the vortex electron wave packet
exhibits a distinctly different deformation under relativistic parameters.
In terms of the probability density distribution in the $xy$ plane,
this deformation manifests as follows: the two parts of the wave packet,
separated by the singularity line, show a tendency to separate from each other
during evolution. Meanwhile, this deformation even alters the direction
of the phase singularity line.
The physical origin of this deformation lies in the time evolution factor
$\mathrm{e}^{-\mathrm{i} \epsilon_{nlsp_z} t}$ of the constituent eigenstates of the wave packet.
To compare the nonrelativistic and relativistic dynamics,
we expand the energy spectrum around the mean occupied
Landau index $\bar N\approx p_Q^2/4$
(see Appendix~\ref{app:choose_range} for details).
For sufficiently small $p_z$ and a narrow distribution
of occupied Landau indices, the expansion gives Eq.~\ref{eq:epsilon_N},
where $N_{nls}=n+\frac{l+|l|+1}{2}+s$ and
$\bar\epsilon=\sqrt{m_e^2+2eB\bar N}$.
We need to distinguish the different physical effects
of the contributions in the $z$ direction and the $xy$ plane:
the kinetic energy term in the $z$ direction,
$\frac{p_z^2}{2\bar\epsilon}$, primarily causes
spreading of the wave packet along the $z$ axis;
whereas the magnetic-field-related terms govern
the relative phases associated with the $xy$ degrees of freedom.
In the nonrelativistic limit, $\bar\epsilon\approx m_e$,
and the quadratic term in $N_{nls}-\bar N$ is negligible.
The energy associated with the $xy$ degrees of freedom
is therefore approximately linear in $N_{nls}$.
This linear relationship implies that after each full
cyclotron period
$T_c\approx T_{\mathrm{NR}}=\frac{2\pi m_e}{eB}$,
the relative phase increments between Landau components
in the $xy$ degrees of freedom are integer multiples of $2\pi$.

Therefore, despite the spreading in the $z$ direction, the wave packet maintains
relatively strict periodic motion in the $xy$ plane, consistent with our previous calculations.
This can be understood by viewing the constituent eigenstates of the wave packet as a set of oscillators with uniform frequency spacings, whose phases realign after one period.
However, in the relativistic case, the quadratic term
in Eq.~\ref{eq:epsilon_N} introduces a nonlinear dependence
of the energy levels on the Landau index.
The nonlinearity introduced by relativity destroys the originally equidistant energy level structure,
causing the relative phases among eigenstate components with different quantum numbers ($n,l,s$)
to fail to restore periodically.
This theoretically accumulated non-uniform phase drift leads to continuous spreading
of the wave packet during evolution, preventing it from periodically regrouping
as in the nonrelativistic case.
In this sense, the constituent eigenstates of the wave packet can be viewed as a set of oscillators with nonuniform frequency spacings, which gradually lose synchronization over time.
At the same time, as shown in the panels (a)-(e) of Fig.~\ref{fig:L_total_R},
influenced by the phase drift, the canonical OAM expectation value of the relativistic
vortex electron significantly deviates from the nonrelativistic result,
especially after one period, where the canonical OAM no longer returns to its initial value. And the deviation of the total canonical OAM from the purple dashed line in Fig.~\ref{fig:L_total_R}(a)-(e) is also reflected in the change in the direction of the phase singularity line in Fig.~\ref{fig:evolution_3}(f)-(g).

\begin{equation}
\label{eq:epsilon_N}
\epsilon_{nlsp_z}\approx
\bar\epsilon+\frac{p_z^2}{2\bar\epsilon}
+\frac{eB}{\bar\epsilon}(N_{nls}-\bar N)
-\frac{(eB)^2}{2\bar\epsilon^3}(N_{nls}-\bar N)^2,
\end{equation}

Further analysis of the relativistic results,
shown by the green and red solid curves in
panels (f) and (g) of Fig.~\ref{fig:L_total_R},
shows that they differ substantially from the
nonrelativistic results and the precession
predicted by the Thomas-BMT-type equation
(gray dashed curves),
which predicts that, in a uniform magnetic
field with the electron velocity perpendicular
to the field, the intrinsic kinetic OAM undergoes
Larmor precession with angular frequency
$ \boldsymbol{\Omega}_L= \frac{e}{2m_e}\frac{2-\gamma}{\gamma}\boldsymbol{B},\  \gamma=\sqrt{1+\frac{\boldsymbol{p}^{\,2}}{m_e^2}}$ \cite{karlovets2026angular}. In particular, the
relativistic OAM components exhibit much larger
oscillation amplitudes that grow markedly over
time, owing to the phase drift discussed above.
From the evolution characteristics of the wave
packet, this phase drift is likely to cause
pronounced deformation of the electron wave
packet and gradual degradation of its vortex
structure. Although this explanation is a physical inference drawn from the numerical results rather than a strict proof, from a practical perspective, long-time evolution of vortex electrons in a transverse magnetic field should be avoided if one aims to preserve a well-defined intrinsic OAM structure.

More quantitatively, the phase drift associated with
the nonequidistant Landau levels can be estimated
from the quadratic term in Eq.~\ref{eq:epsilon_N}.
For negligible $p_z$, the cyclotron period is
$T_c=2\pi\bar\epsilon/(eB)$.
Over one period, the linear contribution produces
relative phase increments that are integer multiples
of $2\pi$, while the quadratic contribution produces
a residual phase mismatch.
For an occupied Landau-index distribution with
characteristic width $\Delta N$, the accumulated
phase drift is estimated as
$\Delta\phi(t)\sim
\frac{(eB)^2}{2\bar\epsilon^3}(\Delta N)^2t
=\frac{\pi eB(\Delta N)^2}{\bar\epsilon^2}
\frac{t}{T_c}$.
Using the estimate
$\Delta N\sim p_{my}/(eB\sigma)$, based on the
Gaussian momentum spread $\Delta p\sim1/\sigma$,
this expression becomes
$\Delta\phi(t)\sim\pi\eta_\phi\,t/T_c$,
where $\eta_\phi=\beta^2/(eB\sigma^2)$ is a dimensionless
parameter and $\beta=p_{my}/\bar\epsilon$.
Thus, suppressing this source of dephasing over
$N_{\mathrm{periods}}=t/T_c$ cyclotron periods
requires $N_{\mathrm{periods}}\pi\eta_\phi\ll1$,
which provides an approximate quantitative criterion
for limiting the phase drift discussed above.

The three examples in Figs.~\ref{fig:evolution_4}(a)-(e), \ref{fig:evolution_4}(f)-(j), and \ref{fig:evolution_2}(f)--(j) provide a quantitative comparison of this phase drift. For the first two cases, with $B=10^5\,\mathrm{T}$ and $\sigma=150\,\mathrm{pm}$, the momenta $p_{my}=2$ and $0.2\,\mathrm{MeV}$ give $\eta_\phi\simeq0.275$ and $0.0389$, respectively. For the nonrelativistic case, with $B=2\,\mathrm{T}$, $p_{my}=0.3\,\mathrm{keV}$, and $\sigma=40\,\mathrm{nm}$, we obtain $\eta_\phi\simeq7.09\times10^{-8}$. The estimated phase drifts over one cyclotron period are therefore $0.86$, $0.12$, and $2.23\times10^{-7}\,\mathrm{rad}$, respectively. The first set shows pronounced elongation and distortion of the density lobes, whereas the second retains a more compact two-lobe profile with weaker deformation; the third provides a nonrelativistic reference for which the estimated relativistic phase drift is negligible.

For an experimentally accessible field $B=1\,\mathrm{T}$, with $\sigma=100\,\mathrm{nm}$ and $p_{my}=2\,\mathrm{MeV}$, we obtain $\eta_\phi\simeq0.0618$, corresponding to an estimated phase drift of $0.194\,\mathrm{rad}$ per cyclotron period, comparable to that in Fig.~\ref{fig:evolution_4}(f)-(j). This estimate highlights the usefulness of $\eta_\phi$, which combines the effects of magnetic field strength, wave-packet width, and electron momentum into a single dimensionless measure of relativistic dephasing. Together with the evolution time, it provides the approximate criterion $N_{\mathrm{periods}}\pi\eta_\phi\ll1$ for limiting the accumulated nonlinear phase mismatch among eigenstate components, and hence a practical guide for assessing dephasing-induced wave-packet distortion and the susceptibility of the vortex structure to degradation during evolution in a transverse magnetic field.

The criterion based on $\eta_\phi$ describes coherent dephasing in a prescribed magnetic field and does not account for radiation effects. To assess this limitation, we estimate the fractional loss of the electron's total energy per cyclotron period using the classical synchrotron-radiation power:
$\eta_{r}=P_{\mathrm{syn}}T_c/(\gamma m_ec^2)
=(4\pi\alpha/3)(B/B_{\mathrm{cr}})\gamma^2\beta_\perp^2$,
where $\alpha$ is the fine-structure constant, $B_{\mathrm{cr}}=m_e^2c^2/(e\hbar)\simeq4.41\times10^9\,\mathrm{T}$, and $\beta_\perp$ is the velocity component perpendicular to the field. The largest estimated loss among the parameter sets considered occurs at $B=4\times10^6\,\mathrm{T}$ and $p_{my}=2\,\mathrm{MeV}$, giving $\eta_{r}\simeq4.2\times10^{-4}$ and a quantum radiation parameter $\chi=\gamma\beta_\perp B/B_{\mathrm{cr}}\simeq3.5\times10^{-3}\ll1$, consistent with the classical estimate. The small fractional loss supports neglecting radiative changes to the mean energy and trajectory over the short evolution times considered, although cumulative radiation effects may become important over longer times. The criterion based on $\eta_\phi$ is derived with radiation neglected; its applicability to vortex-state preservation when photon emission and radiation reaction are included requires further analysis.

\section{\label{sec:conclusion} Conclusion}
The quantum theoretical framework based on the Dirac equation indicates that the dynamics of vortex electrons in a uniform static transverse magnetic field cannot be adequately captured by the rigid point-dipole model.
In the nonrelativistic regime, after one complete cyclotron period, the canonical OAM does not reverse, but instead retains its initial orientation, and its magnitude exhibits intricate fluctuations throughout the evolution. In the symmetric gauge used here, the canonical OAM remains aligned with the phase singularity line throughout the evolution, providing a phase-sensitive characterization of the vortex structure.
The intrinsic kinetic OAM also deviates from simple Larmor precession. The exact equations of motion identify the nonstationary internal current distribution as the origin of this deviation, through additional contributions associated with evolving spatial correlations within the wave packet. The numerical results for both canonical and kinetic OAM are in excellent agreement with the exact analytical solutions, validating our numerical framework.
This quantum behavior highlights the crucial role of the internal wave-packet structure encoded in the wavefunction in governing the dynamical evolution.
Extending the analysis to the relativistic regime, the nonlinear dependence of the energy levels on the quantum numbers breaks the equidistant structure of the energy levels,
causing the relative phases among the eigenstate components constituting the wave packet to drift nonuniformly in time and gradually lose synchrony.
This, in turn, leads to wave-packet spreading and deformation and modifies the intrinsic OAM evolution.
To quantify this effect, we introduce the dimensionless parameter $\eta_\phi=\beta^2/(eB\sigma^2)$, which combines the effects of magnetic field strength, wave-packet width, and electron momentum and estimates the accumulated phase drift as $\Delta\phi(t)\sim\pi\eta_\phi t/T_c$.
In view of these features, for application scenarios that require maintaining the vortex structure and OAM controllability,
the approximate criterion $N_{\mathrm{periods}}\pi\eta_\phi\ll1$, with $N_{\mathrm{periods}}=t/T_c$, provides a quantitative guide for selecting physical parameters and limiting the evolution time to suppress dephasing-induced distortion and mitigate the degradation of vortex characteristics.
Overall, this work not only clarifies the limitations of the rigid point-dipole model in describing vortex electrons, but also provides a quantitative picture of how internal current dynamics and relativistic spectral effects shape the evolution of vortex electron wave packets, which may be useful for future studies and applications of high energy vortex electrons.

\begin{acknowledgments}
We thank Dmitry V. Karlovets, Gregg M. Gallatin, and Igor P. Ivanov for fruitful discussions and valuable suggestions regarding this work. This work is supported by the National Science Foundation of China (Grant No. 12388102), Strategic Priority Research Program of the Chinese Academy of Sciences
(Grant No. XDB0890000), and CAS Project for Young Scientists in Basic Research (Grant No. YSBR060).
\end{acknowledgments}

\appendix
\section{\label{app:solve_dirac} Solutions of the Dirac Equation in a Magnetic Field in Cylindrical Coordinates}
Eq.~\ref{eq:dirac} can alternatively be expressed as:

\begin{equation}
\label{eq:dirac1}
\begin{pmatrix}
i\partial_t - m_e & 0 & i\partial_z & \Delta_B^+ \\
0 & i\partial_t - m_e & \Delta_B^- & -i\partial_z \\
i\partial_z & \Delta_B^+ & i\partial_t + m_e & 0 \\
\Delta_B^- & -i\partial_z & 0 & i\partial_t + m_e
\end{pmatrix}
\begin{pmatrix}
\psi_1 \\ \psi_2 \\ \psi_3 \\ \psi_4
\end{pmatrix}
= 0,
\end{equation}
where $\Delta_B^{\pm}=\mathrm{i} \mathrm{e}^{\mp \mathrm{i} \varphi} \left(\partial_r \mp \frac{\mathrm{i}}{r} \partial_\varphi \pm \frac{eBr}{2} \right)$,
and $\psi_i$ is the $i$-th component ($i=1,2,3,4$) of the wavefunction $\psi(\boldsymbol{x},t)$.
It is observed that the matrix in Eq.~\ref{eq:dirac1} is independent of $\varphi$, $z$, and $t$.
Therefore, we can set the form of the positive energy solution as:
\begin{equation}
\label{eq:psi_i}
\psi_i=\psi_i(r,\varphi,z,t)=C_E \, \Phi_i(r)\, \mathrm{e}^{\mathrm{i} j_i \varphi}\,  \mathrm{e}^{-\mathrm{i}(Et - p_z z)},
\end{equation}
where we take $j_{1,3}=j-\frac{1}{2}$ and $j_{2,4}=j+\frac{1}{2}$.
Substituting Eq.~\ref{eq:psi_i} into Eq.~\ref{eq:dirac1}, we obtain the following system of equations for $\Phi_i(r)$:

\begin{equation}
\label{eq:Phi_part1}
(E\mp m_e) \Phi_{1,3}(r) - p_z \Phi_{3,1}(r) + \mathrm{i}(\partial_r+\frac{j+\frac{1}{2}}{r}+\frac{eBr}{2}) \Phi_{4,2}(r) =0,
\end{equation}

\begin{equation}
\label{eq:Phi_part2}
(E\mp m_e) \Phi_{2,4}(r) + p_z \Phi_{4,2}(r) + \mathrm{i}(\partial_r-\frac{j-\frac{1}{2}}{r}-\frac{eBr}{2}) \Phi_{3,1}(r) =0.
\end{equation}

Applying the operator $(\partial_r+\frac{j+\frac{1}{2}}{r}+\frac{eBr}{2})$ to Eq.~\ref{eq:Phi_part2}
and utilizing Eq.~\ref{eq:Phi_part1} to eliminate $\Phi_{2}$ and $\Phi_{4}$,
we obtain an equation involving only $\Phi_{3,1}$:
\begin{equation}
\label{eq:Phi_13_1}
\left[\eta \partial_\eta^2 +\partial_\eta +\frac{p_\perp^2}{2eB} -\frac{(j-\frac{1}{2})^2}{4\eta} -\frac{j+\frac{1}{2}}{2}-\frac{\eta}{4}\right] \Phi_{3,1}(\eta)=0.
\end{equation}

Similarly, applying $(\partial_r-\frac{j-\frac{1}{2}}{r}-\frac{eBr}{2})$ to Eq.~\ref{eq:Phi_part1}
and utilizing Eq.~\ref{eq:Phi_part2} to eliminate $\Phi_{1}$ and $\Phi_{3}$,
we obtain an equation involving only $\Phi_{4,2}$:
\begin{equation}    
\label{eq:Phi_24_1}
\left[\eta \partial_\eta^2 +\partial_\eta +\frac{p_\perp^2}{2eB} -\frac{(j+\frac{1}{2})^2}{4\eta} -\frac{j-\frac{1}{2}}{2}-\frac{\eta}{4}\right] \Phi_{4,2}(\eta)=0,
\end{equation}
where $p_\perp=\sqrt{E^2-m_e^2-p_z^2}$ and $\eta=\tilde{r}^2=\frac{eB}{2}r^2$.

Investigating the asymptotic behavior of Eqs.~\ref{eq:Phi_13_1} and \ref{eq:Phi_24_1}:
For $\eta\rightarrow\infty$, we have $(\eta\partial_\eta^2 -\frac{\eta}{4})\Phi_{1,2,3,4}=0$, with solutions $\Phi_{1,2,3,4} \rightarrow \mathrm{e}^{-\eta/2}$.
For $\eta\rightarrow 0$, we have $\Big[\eta\partial_\eta^2+\partial_\eta -\frac{(j\mp \frac{1}{2})^2}{4\eta}\Big]\Phi_{\{1,3\},\{2,4\}}=0$,
implying $\Phi_{\{1,3\},\{2,4\}} \rightarrow \eta^{|j\mp \frac{1}{2}|/2}$.
Thus, we can propose the solution in the form:
\begin{equation}
\label{eq:Phi_eta_0}
\begin{aligned}
\Phi_{1,3}(\eta) &= C_{1,3} \eta^{\frac{|j-\frac{1}{2}|}{2}} \mathrm{e}^{-\frac{\eta}{2}}  \Omega_{1}(\eta),  \\
\Phi_{2,4}(\eta) &= C_{2,4} \eta^{\frac{|j+\frac{1}{2}|}{2}} \mathrm{e}^{-\frac{\eta}{2}}  \Omega_{2}(\eta).  
\end{aligned}
\end{equation}

By substituting into Eqs.~\ref{eq:Phi_13_1} and \ref{eq:Phi_24_1}, we derive the equations for $\Omega_{1}(\eta)$ and $\Omega_{2}(\eta)$:
\begin{equation}
\label{eq:Omega_12_0}
\begin{aligned}
&\left[\eta \partial_\eta^2  + (|j-\frac{1}{2}| +1 -\eta) \partial_\eta + n_1\right] \Omega_{1}(\eta) =0,
\\
&\left[\eta \partial_\eta^2  + (|j+\frac{1}{2}| +1 -\eta) \partial_\eta + n_2\right] \Omega_{2}(\eta) =0,    \\
&n_1=\frac{p_\perp^2}{2eB} -\frac{j-\frac{1}{2}+|j-\frac{1}{2}|}{2}-1,\\
&n_2=\frac{p_\perp^2}{2eB} -\frac{j+\frac{1}{2}+|j+\frac{1}{2}|}{2}.
\end{aligned}
\end{equation}

Since the generalized Laguerre polynomials $L_n^{l}(x)$ are solutions to
$\Big[x \frac{\mathrm{d}^2}{\mathrm{d}x^2} +(l+1 -x) \frac{\mathrm{d}}{\mathrm{d}x} + n \Big] y=0$, we obtain:
\begin{equation}
\label{eq:Phi_eta_1}
\begin{aligned}
\Phi_{1,3}(\eta) &= C_{1,3}\, \eta^{\frac{|j-\frac{1}{2}|}{2}} \mathrm{e}^{-\frac{\eta}{2}}  L_{n_1}^{|j-\frac{1}{2}|}(\eta),  \\
\Phi_{2,4}(\eta) &= C_{2,4}\, \eta^{\frac{|j+\frac{1}{2}|}{2}} \mathrm{e}^{-\frac{\eta}{2}}  L_{n_2}^{|j+\frac{1}{2}|}(\eta) . 
\end{aligned}   
\end{equation}

Next, we relate $n_1$ and $n_2$ to the quantum numbers $n$, $l$, and $s$.
With $l=j-s$ ($s=\pm \frac{1}{2}$), and $n$ defined by the energy (including $z$-kinetic, Landau, and Zeeman terms)
$E^2=m^2_e+p_z^2+eB(2n+|l|+l+1+2s)$, or $n=\frac{p_\perp^2}{2eB}-\frac{|l|+l+1}{2}-s$, we obtain:
\begin{equation}
\label{eq:n1_n2}
\begin{aligned}
 \text{When}\  j >0:& \ \ n_1 =n_2= n ,\\
\text{When}\  j <0 \  \text{and}\ l<0,\, s=\frac{1}{2}:& \ \ n_1 = n ,\  n_2=n+2s , \\
 \text{When}\  j <0 \  \text{and}\  l\le0,\, s=-\frac{1}{2}:&\ \  n_1 = n+2s  ,\  n_2=n . \\
\end{aligned}   
\end{equation}

Substituting these results into Eqs.~\ref{eq:Phi_13_1} and \ref{eq:Phi_24_1} determines the relations for coefficients $C_{1,2,3,4}$:

For $j>0$, given $n_1=n_2=n$, we have:
\begin{equation}
\label{eq:C_jg0_eq_1}
\begin{aligned}
C_{4,2} \sqrt{2eB} \left[(j+\frac{1}{2})L^{j+\frac{1}{2}}_n(\eta)-\eta L^{j+\frac{3}{2}}_{n-1}(\eta)\right] &= \mathrm{i}\left[(E\mp m_e)C_{1,3}- p_z C_{3,1}\right] L^{j-\frac{1}{2}}_n(\eta),  
\\
 -C_{3,1} \sqrt{2eB} \left[L^{j-\frac{1}{2}}_n(\eta)+ L^{j+\frac{1}{2}}_{n-1}(\eta)\right] &= \mathrm{i}\left[(E\mp m_e)C_{2,4}+ p_z C_{4,2}\right] L^{j+\frac{1}{2}}_n(\eta).
\end{aligned}
\end{equation}

Using the identities $lL^l_n(x)-xL^{l+1}_{n-1}(x)=(l+n)L^{l-1}_n(x)$ and $L^l_n(x)+L^{l+1}_{n-1}(x)=L_n^{l+1}(x)$,
the above equations simplify to:
\begin{equation}
\label{eq:C_jg0_eq_2}
\begin{aligned}
C_{4,2}\sqrt{2eB} (j+\frac{1}{2}+n) &= \mathrm{i} \left[(E\mp m_e)C_{1,3}- p_z C_{3,1}\right],  \\
 -C_{3,1} \sqrt{2eB} &= \mathrm{i}\left[(E\mp m_e)C_{2,4}+ p_z C_{4,2}\right].
\end{aligned}
\end{equation}
The four equations above provide two effective constraints:
\begin{equation}
\label{eq:C_jg0_3}
\begin{aligned}
\text{When}\ j>0:\ \ &C_3=\frac{p_z}{E+m_e}C_1+ \frac{\mathrm{i}(n+j+\frac{1}{2})}{\sqrt{2eB}(E+m_e)}C_2,\\
&C_4 =  \frac{\mathrm{i}\sqrt{2eB}}{E+m_e} C_1 - \frac{p_z}{E+m_e} C_2.
\end{aligned}   
\end{equation}

For $j<0$, given $n_2=n_1+1$, we obtain similar relations:
\begin{equation}
\label{eq:C_jl0_1}
\begin{aligned}
-C_{4,2}\sqrt{2eB}L^{-j+\frac{1}{2}}_{n_1}(\eta)=\mathrm{i}[(E\mp m_e)C_{1,3}- p_z C_{3,1}] L^{-j+\frac{1}{2}}_{n_1}(\eta),\\
-C_{3,1} \sqrt{2eB}\left[(\eta+j-\frac{1}{2})L^{-j+\frac{1}{2}}_{n_1}(\eta)+\eta L^{-j+\frac{3}{2}}_{n_1-1}(\eta)\right] =\mathrm{i}[(E\mp m_e)C_{2,4}+ p_z C_{4,2}] L^{-j-\frac{1}{2}}_{n_1+1}(\eta).
\end{aligned}
\end{equation}

Using the relation $(x-l)L^{l}_{n}(x)+xL^{l+1}_{n-1}(x)=-(n+1)L^{l-1}_{n+1}(x)$,
this simplifies to two effective constraints:
\begin{equation}
\label{eq:C_jl0_3}
\begin{aligned}
\text{When}\ j<0:\ \ &C_3=\frac{p_z}{E+m_e}C_1+ \frac{\mathrm{i}\sqrt{2eB}}{E+m_e}C_2,\\
&C_4 = - \frac{\mathrm{i}(n_1+1)}{\sqrt{2eB}(E+m_e)} C_1 - \frac{p_z}{E+m_e} C_2.
\end{aligned}  
\end{equation}

With only two constraint equations and thus two remaining degrees of freedom,
we select the independent bases: $(C_1=1, C_2=0)$ and $(C_1=0, C_2=1)$.
In the nonrelativistic and weak field limits, the former
reduces to the eigenstate of the $z$ component spin
operator
$\hat{S}_z=\frac{1}{2}\Sigma_z=\frac{1}{2}\mathrm{diag}(\sigma_z,\sigma_z)$
with eigenvalue $1/2$, while the latter reduces to
the eigenstate with eigenvalue $-1/2$. In more
general cases, however, $\hat{S}_z$ no longer commutes
with the Hamiltonian, so these two solutions are
not exact eigenstates of $\hat{S}_z$ but remain simultaneous eigenstates of $\hat{H}$, $\hat{J}_z$, and $\hat{p}_z$, where $\hat{J}_z=-\mathrm{i}\partial_\varphi+\hat{S}_z$ and $\hat{p}_z=-\mathrm{i}\partial_z$ are the $z$-components of the canonical total angular momentum and canonical momentum, respectively. For convenience, we
still use $s=\pm \tfrac12$ to label these two
linearly independent bases. Consequently, the stationary positive energy cylindrical solutions to the Dirac equation
are presented in Eq.~\ref{eq:psi_pos_energy}.

The negative energy solutions of the Dirac equation
can be obtained by replacing $E$ with $-E$, and the
derivation is essentially the same as that above.
Since the corresponding constraint equations still
provide only two independent conditions, the
negative energy solution space also retains two
degrees of freedom. One may therefore choose two
linearly independent bases to represent the general
solution. In this work, we choose $(C_3=1, C_4=0)$ and $(C_3=0, C_4=1)$ as
the basis states for the negative energy branch to
match the coefficient structure of the standard
negative energy Dirac solutions in the free limit.

In fact, the cylindrical-coordinate solutions of the Dirac equation in a constant magnetic field have a long history of study, and systematic results were already given by Sokolov and Ternov in the 1960s~\cite{sokolov1968synchrotron,sokolov1974relativistic}. The derivation of the Dirac eigenstates adopted in this work mainly follows Ref.~\cite{rajabi2017relativistic}, and has been checked to be consistent with the results of Ref.~\cite{van2017nonuniform}. It should be emphasized that the eigenstates used here and those presented in the book by Sokolov and Ternov essentially correspond to the same solution space of a given Landau level. The difference between them lies mainly in the labeling of the quantum numbers and in the choice of independent basis states within the degenerate subspace. More specifically, after imposing the constraint equations (Eqs.~\eqref{eq:C_jg0_3} and \eqref{eq:C_jl0_3}), only two degrees of freedom remain in the spinor coefficients ($C_{1,2,3,4}$), and the present work and the results of Sokolov and Ternov simply choose different bases in this two-dimensional subspace. The two sets of eigenstates are related by a unitary transformation within the degenerate subspace, and therefore both constitute orthonormal and complete sets of solutions of the Dirac equation.

\section{\label{app:I_xy_and_Q_nlm} Explicit expressions for $I_{xy}^{\pm}$ and evaluation of $Q_{nlm}$}
In the main text, we presented only the representative case of the
$x$-$y$ dependent factor $I_{xy}^{\pm}$ for $j>0$ and $s>0$.
For completeness, we collect here the explicit expressions for the remaining
three sign combinations of $(j,s)$:
\begin{equation}
\label{eq:I_xy}
I_{xy}^{\pm}(n,l,s,p_z,m)=
\begin{cases}
\begin{aligned}
&\left(
\sqrt{E_0 \pm m_e}
\mp \frac{p_z \sqrt{E_0 \mp m_e}}{\epsilon_{nlsp_z}+m_e}
\right)\omega^{\uparrow}_{y,2} Q_{nlm} \\
&\qquad \pm \mathrm{i}\,
\frac{\sqrt{2eB(E_0 \mp m_e)}}{\epsilon_{nlsp_z}+m_e}\,
N_{nls}\,\omega^{\uparrow}_{y,1} Q_{n(l-1)m}
\end{aligned}
& \text{for } j>0,\ s<0, \\[8pt]

\begin{aligned}
&\left(
\sqrt{E_0 \pm m_e}
\pm \frac{p_z \sqrt{E_0 \mp m_e}}{\epsilon_{nlsp_z}+m_e}
\right)\omega^{\uparrow}_{y,1} Q_{nlm} \\
&\qquad \pm \mathrm{i}\,
\frac{\sqrt{2eB(E_0 \mp m_e)}}{\epsilon_{nlsp_z}+m_e}\,
N_{nls}\,\omega^{\uparrow}_{y,2} Q_{(n+1)(l+1)m}
\end{aligned}
& \text{for } j<0,\ s>0, \\[8pt]

\begin{aligned}
&\left(
\sqrt{E_0 \pm m_e}
\mp \frac{p_z \sqrt{E_0 \mp m_e}}{\epsilon_{nlsp_z}+m_e}
\right)\omega^{\uparrow}_{y,2} Q_{nlm} \\
&\qquad \mp \mathrm{i}\,
\frac{\sqrt{2eB(E_0 \mp m_e)}}{\epsilon_{nlsp_z}+m_e}\,
\omega^{\uparrow}_{y,1} Q_{(n-1)(l-1)m}
\end{aligned}
& \text{for } j<0,\ s<0.
\end{cases}
\end{equation}

The expressions above depend on $Q_{nlm}$, which is defined by a complex integral over $x$ and $y$. 
In the following, we provide the details of its evaluation.

\begin{equation}
\label{eq:Q_nlm0}
\begin{aligned}
Q_{nlm}&=\int_{-\infty}^{+\infty} \mathrm{d}x\int_{-\infty}^{+\infty} \mathrm{d}y\ \mathrm{e}^{\mathrm{i}(xp_{mx}+yp_{my})} \mathrm{e}^{-\frac{x^2+y^2}{2\sigma^2}} {\mathcal{J}_n^l}^*(\tilde{r}^2)\\
&=\int_{-\infty}^{+\infty} \mathrm{d}x\int_{-\infty}^{+\infty} \mathrm{d}y\ \mathrm{e}^{\mathrm{i}(xp_{mx}+yp_{my})-\frac{x^2+y^2}{2\sigma^2}} \Big[\frac{eB}{2} (x^2+y^2)\Big]^{\frac{|l|}{2}} \mathrm{e}^{-\frac{eB}{4}(x^2\mathord{+}y^2)-\mathrm{i}l\varphi(x,y)} L^{|l|}_n\Big[\frac{eB}{2}(x^2\mathord{+}y^2)\Big]\\
&=\frac{2}{eB}\int_{0}^{+\infty} \mathrm{d}\tilde{r}\ \mathrm{e}^{-(\frac{1}{eB\sigma^2}+\frac{1}{2}) \tilde{r}^2 } \tilde{r}^{|l|+1} L^{|l|}_n(\tilde{r}^2)   \int_{0}^{2\pi} \mathrm{d}\varphi \ \mathrm{e}^{\mathrm{i}\sqrt{\frac{2}{eB}}(p_{mx}\tilde{r}\cos \varphi+p_{my} \tilde{r} \sin \varphi)-\mathrm{i}l\varphi},\\
\end{aligned}
\end{equation}
where $\tilde{r}=\sqrt{\frac{eB}{2}(x^2+y^2)}$. Transforming the integral from Cartesian to
polar coordinates and integrating over the angle $\varphi$, we obtain:
\begin{equation}
\label{eq:Q_nlm1}
Q_{nlm}=\frac{4\pi}{eB} \mathrm{e}^{\mathrm{i} l \varphi_0} \int_{0}^{+\infty} \ \mathrm{d}\tilde{r}\ \mathrm{e}^{-(\frac{1}{eB\sigma^2}+\frac{1}{2}) \tilde{r}^2 }\ \tilde{r}^{|l|+1}\ L^{|l|}_n(\tilde{r}^2)\ J_l\left[\tilde{r}\sqrt{\frac{2}{eB}(p_{mx}^2\mathord{+}p_{my}^2)}\right],
\end{equation}
where $\varphi_0=\arctan\frac{p_{mx}}{p_{my}}$. The integration over $\varphi$ utilizes
Eq.~\ref{eq:int_varphi1}, the derivation of which is presented below:

\begin{equation}
\label{eq:int_varphi0}
\begin{aligned}
\int_{0}^{2\pi} \mathrm{d} \theta\ \mathrm{e}^{\mathrm{i}(a \cos\theta +b\sin\theta-l\theta)}
&=\int_{0}^{2\pi} \mathrm{d} \theta\ \mathrm{e}^{\mathrm{i}\left[\sqrt{a^2+b^2} \sin (\theta-\theta_0)-l \theta\right]} \ ,\ \theta_0=\arctan \frac{a}{b}
\\
&=\int_{-\pi+\theta_0}^{\pi+\theta_0} \mathrm{d} \theta'\ \mathrm{e}^{\mathrm{i}\left[-\sqrt{a^2+b^2} \sin \theta'-l \theta'+l(\theta_0-\pi)\right]}
\\
&=\int_{-\pi}^{\pi} \mathrm{d} \theta'\ \mathrm{e}^{\mathrm{i}\left[-\sqrt{a^2+b^2} \sin \theta'-l \theta'+l(\theta_0-\pi)\right]}.
\end{aligned}
\end{equation}
The final step of Eq.~\ref{eq:int_varphi0} exploits the periodicity of the integrand
$\mathrm{e}^{\mathrm{i}\left[-\sqrt{a^2+b^2} \sin \theta'-l \theta'+l(\theta_0-\pi)\right]}$.
Due to this periodicity, shifting the integration limits leaves the result invariant.
We then apply Eq.~(8.411.1) from Ref.~\cite{gradshteyn2014table}:
\begin{equation}
\int_{-\pi}^{\pi} \mathrm{d} \theta\ \mathrm{e}^{\mathrm{i}\left(x \sin \theta-l \theta\right)}=2\pi J_{l}(x) .
\end{equation}

Combining this result with the parity properties of the integer-order
Bessel function $J_l(z)$, we finally arrive at:
\begin{equation}
\label{eq:int_varphi1}
\begin{aligned}
\int_{0}^{2\pi} \mathrm{d} \theta\ \mathrm{e}^{\mathrm{i}(a \cos\theta +b\sin\theta-l\theta)}
&=2\pi \mathrm{e}^{\mathrm{i} l\theta_0} (-1)^l J_{l}(-\sqrt{a^2+b^2})
\\
&=2\pi \mathrm{e}^{\mathrm{i} l\theta_0} J_{l}(\sqrt{a^2+b^2}) .
\end{aligned}
\end{equation}

Next, we address the integration of Eq.~\ref{eq:Q_nlm1} with respect to $\tilde{r}$.
For this purpose, we employ Eq.~(7.421.4) from Ref.~\cite{gradshteyn2014table}:
\begin{equation}
\label{eq:int_tilde_r}
\begin{aligned}
\int_{0}^{\infty} \mathrm{d}r \ r^{l+1} \mathrm{e}^{-b r^2} L_n^l(a r^2) J_l(p r) = 2^{-l-1} b^{-l-n-1}(b-a)^n p^l\mathrm{e}^{-\frac{p^2}{4b}} L_n^l\left[\frac{ap^2}{4b(a-b)}\right].
\end{aligned}
\end{equation}

However, Eq.~\ref{eq:int_tilde_r} is not directly applicable to Eq.~\ref{eq:Q_nlm1}, because the order of the
Bessel function $J_l\left[\tilde{r}\sqrt{\frac{2}{eB}(p_{mx}^2\mathord{+}p_{my}^2)}\right]$
is not equal to $|l|$. Utilizing the property of integer-order Bessel functions,
$J_{-l}(x)=(-1)^l J_l(x)$, we rewrite the term as:
\begin{equation}
\label{eq:Bessel_porp}
J_l(x)=(-1)^{\frac{l-|l|}{2}}J_{|l|}(x).
\end{equation}

Finally, after integrating over all coordinates in $Q_{nlm}$, we obtain:

\begin{equation}
\label{eq:Q_nlm}
\begin{aligned}
Q_{nlm}=&\frac{4 \pi}{eB} \mathrm{e}^{\mathrm{i} l \varphi_0} (-1)^{\frac{l-|l|}{2}} 2^{-|l|-1}(\frac{1}{eB\sigma^2}+\frac{1}{2})^{-|l|-n-1} (\frac{1}{eB\sigma^2}-\frac{1}{2})^n  \\
&\times \left[\frac{2}{eB}(p_{mx}^2+p_{my}^2)\right]^{\frac{|l|}{2}}\mathrm{e}^{-\frac{\sigma^2(p_{mx}^2+p_{my}^2)}{2+eB\sigma^2}} L_n^{|l|}\left[\frac{2eB\sigma^4(p_{mx}^2+p_{my}^2)}{e^2B^2\sigma^4-4}\right].
\end{aligned}
\end{equation}

\section{\label{app:choose_range}Choice of Summation and Integration Limits}
After obtaining analytically tractable projection
coefficients, the main difficulty in evaluating the
time-dependent wavefunction in
Eq.~(\ref{eq:mtdfw_puls}) shifts to the treatment of
the sums and integrals. Since numerical calculations
cannot handle infinite sums and integrals directly,
the corresponding summation and integration ranges
must be truncated in a reasonable way. Compared with
the spin quantum number $s$, which only takes the
values $\pm 1/2$, and $p_z$, which is restricted by
the Gaussian distribution in Eq.~(\ref{eq:I_z}), the
main challenge is to choose suitable truncation
ranges for the quantum numbers $n$ and $l$.

\begin{figure}[htbp]
    \centering
    \includegraphics[width=0.5\textwidth]{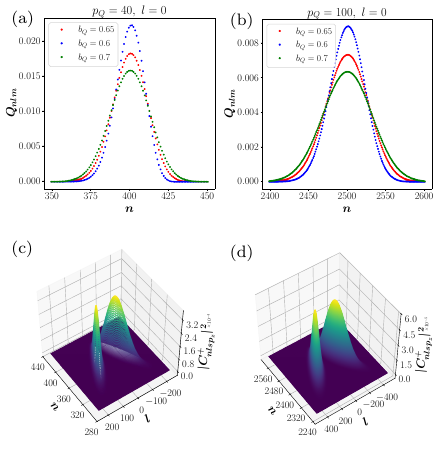}
    \caption{(a) and (b) display the variation of $Q_{nlm}$ with respect to $n$ at $l=0$.
Here, (a) corresponds to $p_Q=40$ and (b) to $p_Q=100$;
the green, red, and blue scattered points represent the cases with $b_Q$ values
of 0.7, 0.65, and 0.6, respectively.
(c) and (d) display the variation of the projection probability $|C_{nlsp_z}^+|^2$
with respect to $n$ and $l$ for $s=1/2$ and $p_z=0$. }  
    \label{fig:nl_range}
\end{figure}

According to Eqs.~\ref{eq:I_xy} and \ref{eq:Q_nlm}, the quantum numbers $n$ and $l$ appear in the exponent of $Q_{nlm}$,
leading to an exponential suppression of $Q_{nlm}$ and subsequently $I_{xy}$.
To explore this effect, we define the parameters $b_Q=\frac{1}{eB\sigma^2}+\frac{1}{2}$ and $p_Q=\sqrt{\frac{2}{eB}(p_{mx}^2+p_{my}^2)}$,
and first analyze the relationship between $Q_{nlm}$ and the quantum number $n$.
Figures~\ref{fig:nl_range}~(a) and (b) display the distribution of $Q_{nlm}$ with respect to $n$ for the case of $l=0$.
Here,  (a) corresponds to $p_Q=40$ and (b) to $p_Q=100$;
the green, red, and blue scattered points represent the cases with $b_Q$ values of 0.7, 0.65, and 0.6, respectively.
Analyzing the figures reveals that $Q_{nlm}$ exhibits an approximately Gaussian distribution with respect to $n$.
The parameter $b_Q$ determines the peak height and width of the distribution, while $p_Q$ determines its center position.

Although the analysis of $Q_{nlm}$ reveals an
exponential suppression with the quantum numbers, it
is still not sufficient to determine the truncation
ranges of $n$ and $l$ directly. To this end, one
must further examine the distribution of the
projection probability $|C^+_{nlsp_z}|^2$ in the
$n,l$ space and determine the summation boundaries
from its main contributing region. 
Figures~\ref{fig:nl_range}(c) and (d) illustrate the distribution of the projection probability $|C_{nlsp_z}^+|^2$ under the conditions $s=1/2$ and $p_z=0$.
Here, the parameter settings for (c) correspond to $p_Q=40$ and $b_Q=0.65$,
while those for (d) correspond to $p_Q=100$ and $b_Q=0.65$.
The figures reveal that the projection probability is mainly
concentrated in a finite region of $n$ and $l$ and
decays rapidly outside this region, which provides a
direct criterion for truncating the sums.

Comprehensive analysis of Fig.~\ref{fig:nl_range} demonstrates a direct correlation
between the parameter $p_Q$ and the value of $n+\frac{l+|l|+1}{2}+s$ at the projection probability peak.
Specifically, $p_Q$ determines the value of $n+\frac{l+|l|+1}{2}+s$
corresponding to the maximum projection probability,
and they approximately satisfy the relationship below:
\begin{equation}
\label{eq:p_Q}
\frac{p_Q^2}{4}=\frac{p_{mx}^2+p_{my}^2}{2eB} \approx \left(n+\frac{l+|l|+1}{2}+s\right)_{\text{most}\ \text{probable}}.
\end{equation}

This relationship is underpinned by a profound physical intuition:
the initial state tends to project onto a basis
with physical properties (such as momentum and energy) similar to its own.
That is, when the $xy$-plane momentum of the eigenstate
$p_{xy}^2=2eB(n+\frac{l+|l|+1}{2}+s)$ matches the
$xy$-plane momentum of the initial state $(p_{mx}^2+p_{my}^2)$,
their wavefunctions overlap maximally, yielding the maximum projection probability.
This is the physical meaning of Eq.~(\ref{eq:p_Q}).
In summary, by selecting a finite range of quantum numbers $\{n,l,s,p_z\}$
centered around the peak of the projection probability,
accurate numerical simulation of the time evolution of the initial state can be achieved.

\section{\label{app:Exact_nonrelativistic_solution} Exact nonrelativistic evolution of the transverse intrinsic kinetic orbital angular momentum}
The nonrelativistic orbital Hamiltonian in a uniform magnetic field is quadratic in position and momentum. Consequently, the equations for the second moments form a closed system, allowing the transverse intrinsic kinetic OAM to be obtained analytically.

We consider an electron of charge $-e$, with $e>0$, and use $\hbar=1$. The magnetic field, kinetic momentum, and orbital Hamiltonian are defined by

\begin{equation}
\boldsymbol B=B\bar{z},\qquad
\boldsymbol\pi=\boldsymbol p+e\boldsymbol A,\qquad
H=\frac{\boldsymbol\pi^2}{2m_e},\qquad
\omega_c=\frac{eB}{m_e},\qquad
\varphi=\omega_ct,
\label{eq:definitions}
\end{equation}

where $B>0$. The spin coupling in a spatially uniform magnetic field does not modify the orbital equations considered here and is omitted.

The intrinsic kinetic OAM is defined using the centered operators,

\begin{equation}
\langle\boldsymbol L_{\mathrm{int}}^{\mathrm{kin}}\rangle
=\langle\delta\boldsymbol r\times\delta\boldsymbol\pi\rangle,
\qquad
\delta\boldsymbol r=\boldsymbol r-\langle\boldsymbol r\rangle,\qquad
\delta\boldsymbol\pi=\boldsymbol\pi-\langle\boldsymbol\pi\rangle.
\label{eq:intrinsic_oam}
\end{equation}

Since the equations of motion are linear, the centered operators obey the same equations as the uncentered ones. Below, the symbols $\delta$ are suppressed: all moments are centered moments, and the OAM components refer to intrinsic quantities.

For an initial vortex oriented along the $y$ axis, the relevant initial conditions are

\begin{equation}
\begin{gathered}
\langle z\pi_x\rangle_0=-\langle x\pi_z\rangle_0=\frac{l_0}{2},
\qquad
\langle\pi_z\pi_y\rangle_0=c,\qquad
\kappa=\frac{c}{m\omega_c},\\
\langle zx\rangle_0=\langle zy\rangle_0
=\langle z\pi_y\rangle_0=\langle y\pi_z\rangle_0
=\langle\pi_z\pi_x\rangle_0=0.
\end{gathered}
\label{eq:initial_conditions}
\end{equation}

Thus, the initial transverse kinetic OAM is
$\langle\boldsymbol L_\perp^{\mathrm{kin}}\rangle_0=(0,l_0)$, where
$\boldsymbol L_\perp^{\mathrm{kin}}=(L_x^{\mathrm{kin}},L_y^{\mathrm{kin}})$.

The Heisenberg equations,
$\dot{\boldsymbol r}=\boldsymbol\pi/m_e$ and
$\dot{\boldsymbol\pi}=\omega_c(-\pi_y,\pi_x,0)$, give

\begin{equation}
\begin{aligned}
\pi_x(t)&=\pi_{x0}\cos\varphi-\pi_{y0}\sin\varphi,
&
\pi_y(t)&=\pi_{x0}\sin\varphi+\pi_{y0}\cos\varphi,\\
x(t)&=x_0+\frac{\pi_{x0}\sin\varphi-\pi_{y0}(1-\cos\varphi)}{m_e\omega_c},
&
y(t)&=y_0+\frac{\pi_{x0}(1-\cos\varphi)+\pi_{y0}\sin\varphi}{m_e\omega_c},\\
z(t)&=z_0+\frac{t}{m_e}\pi_{z0}.
\end{aligned}
\label{eq:heisenberg_solution}
\end{equation}

Using Eq.~\ref{eq:initial_conditions}, the mixed moments entering the transverse OAM become

\begin{equation}
\begin{aligned}
\langle z\pi_x\rangle_t&=\frac{l_0}{2}\cos\varphi-\kappa\varphi\sin\varphi,
&
\langle z\pi_y\rangle_t&=\frac{l_0}{2}\sin\varphi+\kappa\varphi\cos\varphi,\\
\langle x\pi_z\rangle_t&=-\frac{l_0}{2}-\kappa(1-\cos\varphi),
&
\langle y\pi_z\rangle_t&=\kappa\sin\varphi.
\end{aligned}
\label{eq:mixed_moments}
\end{equation}

Substitution into $L_x^{\mathrm{kin}}=y\pi_z-z\pi_y$ and
$L_y^{\mathrm{kin}}=z\pi_x-x\pi_z$ yields the exact solution

\begin{equation}
\langle\boldsymbol L_\perp^{\mathrm{kin}}\rangle_t
=
\left(\frac{l_0}{2}+\kappa\right)\bar{ y}
+\left(\frac{l_0}{2}-\kappa\right)(-\sin\varphi,\cos\varphi)
-\kappa\varphi(\cos\varphi,\sin\varphi).
\label{eq:exact_solution}
\end{equation}

The first term is fixed in the laboratory frame, the second rotates at the cyclotron frequency, and the third is a secular contribution along the rotating radial direction. The coefficient of the secular term is determined entirely by the initial $z,y$ kinetic-momentum correlation $c$.

For the axially symmetric vortex initial state specified in the symmetric gauge, the canonical moments satisfy
$\langle p_zp_y\rangle_0=0$ and
$\langle xp_z\rangle_0=-l_0/2$. Therefore,

\begin{equation}
\boldsymbol A=\frac{B}{2}(-y,x,0),\qquad
c=\left\langle p_z\left(p_y+\frac{eB}{2}x\right)\right\rangle_0
=-\frac{eBl_0}{4},\qquad
\kappa=-\frac{l_0}{4}.
\label{eq:symmetric_gauge}
\end{equation}

Eq.~\ref{eq:exact_solution} then gives

\begin{equation}
\langle L_x^{\mathrm{kin}}\rangle_t
=\frac{l_0}{4}(\varphi\cos\varphi-3\sin\varphi),
\qquad
\langle L_y^{\mathrm{kin}}\rangle_t
=\frac{l_0}{4}(1+3\cos\varphi+\varphi\sin\varphi).
\label{eq:kinetic_oam_components}
\end{equation}

After one cyclotron period,

\begin{equation}
T_c=\frac{2\pi}{\omega_c},\qquad
\langle\boldsymbol L_\perp^{\mathrm{kin}}(T_c)\rangle
=l_0\left(\frac{\pi}{2},1\right).
\label{eq:after_cyclotron_period}
\end{equation}

Hence, the transverse kinetic OAM does not recover its initial orientation after one cyclotron period.

In the symmetric gauge, the canonical and kinetic OAM are related by
$L_x^{\mathrm{kin}}=L_x-eBzx/2$ and
$L_y^{\mathrm{kin}}=L_y-eBzy/2$. For the same initial state, the transverse canonical OAM is

\begin{equation}
\langle L_x\rangle_t
=l_0\left[-\frac12\sin\varphi-\frac{\varphi}{8}(1-\cos\varphi)\right],\quad
\langle L_y\rangle_t
=l_0\left[\frac12(1+\cos\varphi)+\frac{\varphi}{8}\sin\varphi\right].
\label{eq:canonical_oam}
\end{equation}

Although these canonical components return to their initial values at $T_c$, their recurrence does not imply a recurrence of the kinetic OAM.

The gauge-invariant departure from Larmor precession follows directly from the equations of motion:

\begin{equation}
\frac{d}{dt}\langle L_x^{\mathrm{kin}}\rangle
=-\frac{\omega_c}{2}\langle L_y^{\mathrm{kin}}\rangle
-\frac{eB}{2}\frac{d}{dt}\langle zx\rangle,\quad
\frac{d}{dt}\langle L_y^{\mathrm{kin}}\rangle
=\frac{\omega_c}{2}\langle L_x^{\mathrm{kin}}\rangle
-\frac{eB}{2}\frac{d}{dt}\langle zy\rangle,
\label{eq:gauge_invariant_equations}
\end{equation}

with

\begin{equation}
m\frac{d}{dt}\langle zx\rangle=\langle z\pi_x+x\pi_z\rangle,
\qquad
m\frac{d}{dt}\langle zy\rangle=\langle z\pi_y+y\pi_z\rangle.
\label{eq:mixed_spatial_derivatives}
\end{equation}

Thus, the additional torque is governed by the time derivatives of the mixed spatial moments, rather than by the moments themselves. These gauge-invariant quantities characterize the nonstationarity of the internal current distribution. If both mixed moments are stationary, Eq.~\ref{eq:gauge_invariant_equations} reduces to pure Larmor precession. This nonstationarity also gives rise to the tilted two-lobed structure shown in Fig.~\ref{fig:evolution_2}(k)-(o).

The secular behavior also depends on the physical initial state. In particular, for $c=0$,

\begin{equation}
\langle\boldsymbol L_\perp^{\mathrm{kin}}\rangle_t
=\frac{l_0}{2}(-\sin\varphi,\,1+\cos\varphi),
\label{eq:c_zero_case}
\end{equation}

which is periodic and bounded. The secular contribution is therefore not a universal feature of vortex evolution in a uniform magnetic field, but depends on the physical preparation of the initial state. For a fixed physical state, all results are invariant under a simultaneous gauge transformation of the vector potential and wavefunction. Keeping the same wavefunction while changing the vector potential, without the corresponding wavefunction transformation, generally specifies a different physical state and can change $c$. One idealized preparation of the initial state considered here is a sudden switch-on of the magnetic field in the region occupied by an initially free vortex wave packet. In the symmetric gauge, this protocol is described by
\begin{equation}
\boldsymbol A_{\mathrm S}(\boldsymbol r,t)
=f(t)\frac{B}{2}(-y,x,0),
\qquad
\Phi_{\mathrm S}=0,
\label{eq:AS}
\end{equation}
where $f(t)$ increases from zero to unity. The associated physical fields are
$\boldsymbol B(t)=f(t)B\bar{z},\ \boldsymbol E(t)=\frac{B\dot f(t)}{2}(y,-x,0)$.
In the sudden-switching limit, the wavefunction remains approximately unchanged in this gauge, while the kinetic momentum changes through the vector potential. For an incoming free vortex satisfying
$\langle p_zp_y\rangle_{\mathrm{in}}=0$ and
$\langle xp_z\rangle_{\mathrm{in}}=-l_0/2$, the resulting correlation is
$c=\left\langle p_z\left(p_y+\frac{eB}{2}x\right)\right\rangle_{\mathrm{in}}=-\frac{eBl_0}{4}$.

Thus, the value of $c$ used above has a physical preparation interpretation: it results from the kinetic-momentum change induced during the switch-on.

By contrast, a free vortex entering an already established magnetic-field region represents a different preparation protocol and requires a separate treatment of the entry dynamics. For example, an idealized static boundary for incidence along $+y$ can be described by
\begin{equation}
\boldsymbol B(y)=B\Theta(y)\bar{z},
\qquad
\boldsymbol A_{\mathrm L}(y)=-By\Theta(y)\bar{x},
\qquad
\Phi_{\mathrm L}=0.
\label{eq:AL}
\end{equation}

Here the vector potential vanishes in the field-free region and is continuous at the boundary. Taking the same free incoming vortex state as in the switch-on protocol, and neglecting changes to the wave packet during entry, we obtain $c=\langle\pi_z\pi_y\rangle_0
=\langle p_zp_y\rangle_{\mathrm{in}}=0,$
since $\pi_y=p_y$ and $\pi_z=p_z$ in this gauge. Provided that the remaining initial moments satisfy Eq.~\ref{eq:initial_conditions}, the subsequent evolution is given by the bounded, periodic solution in Eq.~\ref{eq:c_zero_case}. Thus, within these idealized preparation models, the same free incoming wave packet leads to different initial kinetic-momentum correlations: $c=-eBl_0/4$ for sudden switch-on and $c=0$ for entry into a preexisting field. This difference reflects the preparation protocol, rather than a gauge transformation of the same physical state. Finally, the exact solution depends only on the initial moments specified above. Any wave packet satisfying these conditions exhibits the same transverse intrinsic kinetic OAM evolution, irrespective of the remaining details of its envelope.

\section{\label{app:Laguerre} Calculation of High-Order Laguerre Polynomials}
To address the inefficiency of iterative algorithms for high-order Laguerre polynomials,
we aim for an efficient, accurate algorithm via high-order asymptotic expansions.
As direct expansions of $L_n^l(x)$ are highly restricted,
we leverage the asymptotic expansion of the Whittaker confluent hypergeometric function $M_{\kappa,\mu}(x)$
to indirectly establish a more robust expansion for $L_n^l(x)$.

Based on Eq.~(18.11.2) in Ref.~\cite{olver2010nist}, the relationship between
the Laguerre polynomial and the Whittaker confluent hypergeometric function is:
\begin{equation}
\label{eq:L_to_M}
 L_n^l(x)=\frac{(l+1)_n}{n!}x^{-\frac{1}{2}(l+1)}e^{\frac{1}{2} x} 
 M_{n+\frac{1}{2}(l+1),\frac{1}{2}l}(x),   
\end{equation}
here, the Pochhammer symbol is defined as $(l+1)_n=\frac{\Gamma(l+n+1)}{\Gamma(l+1)}$.
For the case under study, where $n+\frac{1}{2}(l+1)=\kappa \rightarrow \infty$
and $0\le \frac{1}{2}l=\mu \le (1-\delta) \kappa$, we can directly refer to
Secs.~13.21(ii) and (iii) in Ref.~\cite{olver2010nist} for the asymptotic
expansions of the Whittaker confluent hypergeometric function $M_{\kappa,\mu}(x)$.
Furthermore, the asymptotic form of $M_{\kappa,\mu}(x)$ involves Bessel functions
of the first kind. Since high-order Bessel functions also significantly affect
computational efficiency, we substitute them using the asymptotic expansions of
Bessel functions from Sec.~10.20(i) of Ref.~\cite{olver2010nist}.

\nocite{*}

\bibliography{paper}

\end{document}